\documentclass[twocolumn,dvipdfmx,showpacs,preprintnumbers,amsmath,amssymb,aps,prd,nofootinbib,superscriptaddress,eqsecnum]{revtex4}

\usepackage{makeidx}
\usepackage[dvipdfmx]{graphicx}
\usepackage[all]{xy}
\usepackage{feynmp}
\usepackage{amsmath,amssymb,bm}
\usepackage[usenames]{color}

\usepackage{ulem} 

\allowdisplaybreaks[1]

\newcommand{\todaye}{\the\year /\the\month /\the\day}

\newcommand{\mc}{\mathcal}
\newcommand{\bee}[1]{\begin{align} #1 \end{align}}
\newcommand{\mt}[1]{\begin{matrix} #1 \end{matrix}}

\newcommand{\nn}{\nonumber}
\newcommand{\ra}{\rightarrow}

\newcommand{\Ra}{\Rightarrow}

\newcommand{\rax}[2]{\xrightarrow[#2]{#1}}

\newcommand{\tr}{\mathrm{Tr}}
\newcommand{\pd}{\partial}

\newcommand{\U}[1]{U$(1)_{#1}$}

\newcommand{\braa}[1]{\left( #1 \right)}
\newcommand{\brab}[1]{\left\{ #1 \right\}}
\newcommand{\brac}[1]{\left[ #1 \right]}
\newcommand{\brad}[1]{\left| #1 \right|}
\newcommand{\brae}[1]{\left\langle #1 \right\rangle}
\newcommand{\bra}[1]{\left\langle #1 \right|}
\newcommand{\ket}[1]{\left| #1 \right\rangle}

\begin{document}

\title{
Relations among pionic decays of spin-1 mesons \\
from an SU(4)$\times$U(1) emergent symmetry in QCD}

\author{Hiroki Nishihara\footnote{h248ra@hken.phys.nagoya-u.ac.jp}}
\author{Masayasu Harada\footnote{harada@hken.phys.nagoya-u.ac.jp}}
\affiliation{
Department of Physics, Nagoya University, Nagoya 464-8602, Japan
}

\date{\today}

\def\theequation{\thesection.\arabic{equation}}

\renewcommand{\thefootnote}{\#\arabic{footnote}}

\begin{abstract}
Motivated by recent results by lattice analysis, we assume that the spin-1 mesons of $\left( \rho, \omega, a_1, \rho', \omega', b_1, f_1, h_1\right)$ make a representation of $\mathbf{16}$ of U(4) emergent symmetry in two-flavor QCD when the chiral symmetry is not broken. We study the decay properties of the spin-1 mesons by using a chiral model with an SU(4)$\times$U(1) hidden local symmetry. We first show that, since the SU(4) symmetry is spontaneously broken together with the chiral symmetry, each coupling of the interaction among one pion and two spin-1 mesons is proportional to the mass difference of the relevant spin-1 mesons similarly to the Goldberger-Treiman relation. In addition, some of one-pion couplings are related with each other by the SU(4) symmetry. We further show that there is a relation among the mass of $\rho'$ meson, the $\rho'\pi\pi$ coupling and the $\rho'$-photon mixing strength as well as the Kawarabayashi-Suzuki-Riazuddin-Fayyazuddin relation for the $\rho$ meson. From the relations, we give numerical predictions such as ratios of the spin-1 meson decay widths, which are compared with future experiments for testing the existence of the U(4) emergent symmetry.

\end{abstract}

\pacs{
11.30.Rd,\ 
14.40.Be,\ 
13.75.Lb,\ 
12.39.Fe
}

\maketitle

\section{Introduction}
\label{sec:intro}
\begin{figure}[h]
\centering	\includegraphics[width=80mm]{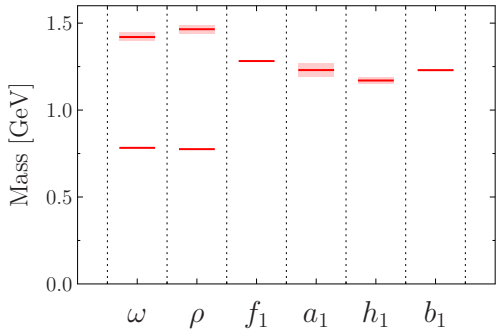}
\caption{Spectra of spin-1 mesons \cite{Agashe:2014kda}.}
\label{Spectra}
\end{figure}

In Quantum ChromeDynamics (QCD) 
the chiral symmetry is one of the most important symmetry to investigate properties of hadron.
In particular, pion is identified as the pseudo Nambu-Goldstone (NG) boson corresponding to the spontaneously symmetry breaking of the chiral symmetry, which means that dynamics of pion is described by the low energy theorem.
On the other hand, there are the spin-1 mesons at about 1 GeV mass, as shown in FIG.\,\ref{Spectra}.
It is a variable problem to describe the spin-1 mesons and pion in a discussion of symmetries and their breaking.

Recently, the existence of an SU$(4)$ symmetry in the spin-1 meson sector is suggested by 
Refs.\,\cite{Denissenya:2014ywa,Glozman:2015qva,Cohen:2015ekf} via the lattice QCD  calculation.  This is also  shown in unbroken limit of the chiral symmetry in two-flavor  case  in Ref.\,\cite{Shifman:2016efc}. 
In the references, the symmetry is called as an emergent symmetry in QCD.
The SU(4) symmetry includes the chiral symmetry SU(2)$_L\times$SU(2)$_R\times$U(1)$_A$, which
is corresponding to the rotation of
the quark field $\psi$ written as $\psi^T=\braa{u_R , d_R , u_L , d_L}$. 
This means that the mesons denoted by $\braa{\rho, a_1, \rho', \omega', b_1, f_1, h_1}$ are members belonging to a multiplet of the SU(4) symmetry,
and that the mass differences  of members  are caused by the spontaneous chiral symmetry breaking.
This indicates that there exist extended Goldberger-Treiman\,(GT)  relations between the mass differences of spin-1 mesons and their couplings to  pions which are Nambu-Goldstone bosons  associated with the chiral symmetry breaking.

In this work, we construct an effective Lagrangian with an $\mbox{SU(4)}\times\mbox{U(1)}$ hidden local symmetry\,(HLS)\,\cite{Bando:1984ej,Bando:1984pw,Bando:1985rf,Bando:1987br,Bando:1987ym,Harada:2003jx},
which includes
the spin-1 mesons, $\braa{\omega, \rho, a_1, \rho', \omega', b_1, f_1, h_1}$, as gauge fields of the HLS. 
The symmetry of the Lagrangian is $\left[ \mbox{SU(2)}_R \times \mbox{SU(2)}_L \times \mbox{U(1)}_A \right]_{\rm chiral} \times \left[ \mbox{SU(4)} \times \mbox{U(1)} \right]_{\rm HLS}$, which is broken to $\mbox{SU(2)}_{\rm isospin}$ symmetry by the chiral condensate.
Then, we show extended GT relations, by which each coupling of the interaction among one pion and two spin-1 mesons is proportional to the mass difference of the relevant spin-1 mesons.  
In addition,  we can derive relations among the coupling of the spin-1 mesons to one pion thanks to the existence of the SU(4) symmetry. 
Furthermore, we show that there is a relation among the mass of $\rho'$ meson, the $\rho'\pi\pi$ coupling and the $\rho'$-photon mixing strength as well as the Kawarabayashi-Suzuki-Riazuddin-Fayyazuddin relations for the $\rho$ meson. 
These relations give us predictions for one-pion decay widths of spin-1 mesons and the electromagnetic form factor of the pion,  which can be verified by future experiments.

In this paper, we conduct the analyses: 
In Sec.\,\ref{sec:construction}, 
we construct  a  Lagrangian with the SU(4)$\times$U$(1)$ HLS to introduce the spin-1 mesons.
In Sec.\,\ref{sec:Vector mesons and pion}, we give eigenstates and  masses of the spin-1 mesons. 
In Sec.\,\ref{sec:GT},  we obtain extended GT relations. We study one-pion decays of  the  spin-1 mesons in section~\ref{sec:one pion decays} and  extended KSRF  relations   in section~\ref{sec:KSRF}.  
In Sec.\,\ref{sec: numerical analysis}, we make a numerical analysis to determine the parameters and give a prediction on the electromagnetic form factor of pion. 
The summery and discussions are given in Sec.\,\ref{sec:summary}.

\section{Construction }
\label{sec:construction}

We construct a chiral Lagrangian with the ${{\rm SU(4)}\times {\rm U(1)}}$ hidden local symmetry.
The Lagrangian has 
the chiral  symmetry~\footnote{
 Explicit breaking of $\mbox{U(1)}_A$ by anomaly is added later together with the  explicit chiral symmetry breaking  from the current quark masses of up and down quarks. }
$\mbox{SU(2)}_{R} \times \mbox{SU(2)}_L \times \mbox{U(1)}_A $ 
and
the gauge symmetry $G_{\rm local}=\brac{\mbox{SU(4)}\times \mbox{U(1)}}_{\rm HLS}$.
Here 
spin-1 mesons are introduced as the gauge fields of $G_{\rm local}$,
which are identified as $\braa{\omega, \rho, a_1, \rho', \omega', b_1, f_1, h_1}$ mesons.
The spontaneously symmetry breaking is represented as
$G_{\rm global} \times G_{\rm local}( =\left[ \mbox{SU(2)}_R \times \mbox{SU(2)}_L \times \mbox{U(1)}_A \right]_{\rm chiral} \times \brac{\mbox{SU(4)}\times \mbox{U(1)}}_{\rm HLS} ) \ra H ( = \mbox{SU(2)}_{V} )$, 
where the NG-bosons identified as the pions and  the  eta meson emerge.

The NG bosons associated with 
the coset-space $G_{\rm global}/H$ are introduced 
through the 2 by 2 unitary matrix field $U$ 
as
\begin{equation}
U = \exp \left( i \frac{\eta}{ f_\eta } + i \sum_{a=1}^3 \frac{ \pi^a \sigma^a }{ f_\pi } \right) \ , 
\end{equation}
where $\eta$ and $\pi^a$ ($a=1,2,3$) are the eta meson and the pion fields, and $\sigma^a$ is the Pauli matrix.
For introducing the $\mbox{SU(4)} \times \mbox{U(1)}$ HLS, we embed this $U$ into 4 by 4 matrix field ${\mathcal U}$ as
\bee{
\mc{U} =
\braa{\mt{
0&U^\dagger\\U &0
}}
}
which transforms under the chiral symmetry $\mbox{SU(2)}_L\times\mbox{SU(2)}_R\times\mbox{U(1)}_A$ as
\begin{equation}
{\mathcal U} \ \to \ {\mathcal G} \cdot {\mathcal U} \cdot {\mathcal G}^\dag \ ,
\end{equation}
where $\mc{G}$ is an element of $G_{\rm global}= \mbox{SU(2)}_L\times\mbox{SU(2)}_R\times\mbox{U(1)}_A$ written as
\bee{
\mc{G} =
\braa{\mt{g_R &0\\0&g_L }}
\braa{\mt{g_A&0\\0&g_A^\dagger}}
}
by using $g_{L,R} \in \mbox{SU(2)}_{L,R}$ and $g_{A} \in \mbox{U(1)}_{A}$.
The generators of the chiral symmetry $G_{\rm global}$ are 
\bee{
T^A_{\rm global}=&
\brab{ 
\frac{S^a + X_{(3)}^a}{\sqrt{2}} \ ,~~
\frac{S^a - X_{(3)}^a}{\sqrt{2}} \ ,~~
X_{(3)}^0
}
\nn\\
=&
\brab{
T^a_{R}
 \ ,~~
T^a_{L}
\ ,~~
X_{(3)}^0
}
\ ,
\label{generators of chiral}
}
whose 
explicit form are given in Appendix~\ref{sec:Generators}.
After the spontaneous symmetry breaking of $G_{\rm global}$,
the generators expressed by $S$ are corresponding to the unbroken ones, while $X$ are broken generators.
This implies that transformations  generated by $S$  belong  to $H$.

Let us decompose $\mc{U}$ as
\bee{
\mc{U} 
=&
\Xi^\dagger (x) 
\cdot
\Xi^\dagger_m (x) 
\cdot
\bar{\Sigma}
\cdot
\Xi_{m} (x) 
\cdot
 \Xi (x)
}
by using 
\bee{
\bar{\Sigma}
\equiv
\braa{\mt{
0&1_2\\1_2&0
}}
\ .
}
These fields transform under $G_{\rm global} \times G_{\rm local}\times H_{\rm extra}$ as
\bee{
\Xi (x) \ra& \tilde{\mc{G}} (x) \cdot \Xi  (x) \cdot \mc{G}^\dagger
\ ,
\nn\\
\Xi_{m} (x)  \ra& \tilde{h} (x) \cdot \Xi_{m} (x) \cdot \tilde{\mc{G}}^\dagger (x)
}
where ${\mc{G}} \in G_{\rm global}$, $\tilde{\mc{G}} (x) \in G_{\rm local}$, and 
$\tilde{h} (x) $ is an element of an $H_{\rm extra} \left( =\mbox{U(2)} \right)$  extra local symmetry,
whose generator is written as
\bee{
T^A_{\rm extra}
=&
\brab{
S^a , S^0 }
\ .
}
From the above transformation properties, 
the covariant derivatives are expressed as
\begin{align}
 D_\mu  \Xi (x) = &
\partial_\mu \Xi(x)  - i V_\mu \Xi(x) +  i \Xi(x) {\mathcal V}_\mu
\notag\\
D_\mu \Xi_{m}(x) = & \partial_\mu \Xi_m(x) - i \tilde{V}_\mu \Xi_m(x) + i \Xi_m(x) V_\mu
\ .
\end{align}
where
$V_\mu$ is the HLS gauge field, $V_\mu = V_\mu^A T^A$,  for $G_{\rm local}$,
$\mc{V}_\mu$ is the external gauge field written by
\bee{
\mc{V}_\mu = \mc{R}_\mu^a \cdot T_R^a +\mc{L}_\mu^a \cdot T_L^a 
+ \sqrt{2} \mc{V}_\mu^0 S^0 +\sqrt{2} \mc{A}_\mu^0 X_{\perp (3)}^0
\ ,
\label{definition of external gauge}
}
and 
$\tilde{V}_\mu$ is the gauge field for $H_{\rm extra}$.
We note that we do not introduce the kinetic term for this $\tilde{V}_\mu$, so that it is not a dynamical field in the present analysis.
The field strength of the HLS gauge field is written as
\bee{
V_{\mu\nu} \equiv  \pd_\mu V_\nu - \pd_\nu V_\mu -i \brac{ V_\mu , V_\nu} \ .
}

For constructing the Lagrangian, it is convenient to introduce the following
covariantized Maurer-Cartan 1-forms:
\bee{
\hat{\alpha}_{\mu} (x) \equiv&
\frac{1}{i}
\Xi_{m} (x) \cdot
\braa{
 D_\mu  \Xi (x) \cdot \Xi^\dagger(x)  
}
\cdot \Xi_{m}^\dagger (x) 
\ ,
\nn\\
\hat{\alpha}_{\mu}^{(m)} (x) \equiv&
\frac{1}{i}
D_\mu \Xi_{m}(x) \cdot \Xi_{m}^\dagger (x) 
\ ,
\label{definition of alpha}
}
which 
transform under $G_{\rm global}\times G_{\rm local}\times H_{\rm extra}$, the parity $\mc{P}$, and the charge conjugation $\mc{C}$ as
in Table\,\ref{Table:transformation properties}.
\begin{table}
\caption[]{Transformation properties of the Maurer-Cartan  1-forms  $\hat{\alpha}_\mu$  and $\hat{\alpha}^{(m)}_\mu$  under $G_{\rm global} \times G_{\rm local}$, $\mc{P}$, and $\mc{C}$ transformations.}
\begin{center}
\begin{tabular}{c|rl}\hline\hline
$G_{\rm global} \times G_{\rm local}$&
$\hat{\alpha}_{\mu} (x)
\ra$&$
\tilde{h} (x) \cdot 
\hat{\alpha}_{\mu} (x)
\cdot 
\tilde{h}^\dagger (x)
$\\
$\mc{P}$ trans.&
$
\hat{\alpha}_{\mu } (x)
\rax{\mc{P}}{}$&$
\bar{\Sigma} \cdot 
\hat{\alpha}_{\mu } (x)
\cdot 
\bar{\Sigma}
$\\
$\mc{C}$ trans.&
$
\hat{\alpha}_{\mu} (x)
\rax{\mc{C}}{}$&$
-\bar{\Sigma} \cdot 
\braa{\hat{\alpha}_{\mu} (x)}^*
\cdot 
\bar{\Sigma}
$\\\hline
$G_{\rm global} \times G_{\rm local}$&
$
\hat{\alpha}_{\mu}^{(m)} (x)
\ra$&$
\tilde{h} (x) \cdot 
\hat{\alpha}_{\mu}^{(m)} (x)
\cdot 
\tilde{h}^\dagger (x)
$\\
$\mc{P}$ trans.&
$
\hat{\alpha}_{\mu }^{(m)} (x)
\rax{\mc{P}}{}$&$
\bar{\Sigma} \cdot 
\hat{\alpha}_{\mu }^{(m)} (x)
\cdot 
\bar{\Sigma}
$\\
$\mc{C}$ trans.&
$
\hat{\alpha}_{\mu}^{(m)} (x)
\rax{\mc{C}}{}$&$
-\bar{\Sigma} \cdot 
\braa{\hat{\alpha}_{\mu}^{(m)} (x)}^*
\cdot 
\bar{\Sigma}
$
\\\hline\hline
\end{tabular}
\end{center}
\label{Table:transformation properties}
\end{table}

Now, the 1-form $\hat{\alpha}_\mu$ is 
classified as
\bee{
\hat{\alpha}_{\mu \parallel } (x)
\equiv&
2\tr\brac{\hat{\alpha}_{\mu} \cdot S^a
}
S^a
\ ,
\nn\\
\hat{\alpha}_{\mu \perp (1)} (x)
\equiv&
2\tr\brac{\hat{\alpha}_{\mu} \cdot X_{(1)}^a
}
X_{(1)}^a
\ ,
\nn\\
\hat{\alpha}_{\mu \perp (2)} (x)
\equiv&
2\tr\brac{\hat{\alpha}_{\mu} \cdot X_{(2)}^a
}
X_{(2)}^a
\ ,
\nn\\
\hat{\alpha}_{\mu \perp (3)}(x)
\equiv&
2\tr\brac{\hat{\alpha}_{\mu} \cdot X_{(3)}^a
}
X_{(3)}^a
\ ,
\nn\\
\hat{\alpha}_{\mu \parallel }^{ (I=0)} (x)
\equiv&
2\tr\brac{\hat{\alpha}_{\mu} \cdot S^0
}
S^0
\ ,
\nn\\
\hat{\alpha}_{\mu \perp (1)}^{ (I=0)} (x)
\equiv&
2\tr\brac{\hat{\alpha}_{\mu} \cdot X_{(1)}^0
}
X_{(1)}^0
\ ,
\nn\\
\hat{\alpha}_{\mu \perp (2)}^{ (I=0)} (x)
\equiv&
2\tr\brac{\hat{\alpha}_{\mu} \cdot X_{(2)}^0
}
X_{(2)}^0
\ ,
\nn\\
\hat{\alpha}_{\mu \perp (3)}^{ (I=0)}(x)
\equiv&
2\tr\brac{\hat{\alpha}_{\mu} \cdot X_{(3)}^0
}
X_{(3)}^0
\ ,
}
and similarly for $\alpha_{\mu}^{(m)}$. 
Note that the $\tilde{V}_\mu$ gauge field for $H_{\rm extra}$ is included only in $\hat{\alpha}_{\mu\parallel}^{(m)}$ and $\hat{\alpha}_{\mu\parallel}^{(m)(I=0)}$.
Furthermore, from 
the definitions \eqref{definition of alpha}
of $\hat{\alpha}_\mu$ and $\hat{\alpha}^{(m)}_\mu$, 
we find that the sum 
\bee{
\hat{\alpha}_\mu (x)+\hat{\alpha}_\mu^{(m)} (x)
=&
\frac{1}{i}
 D_\mu  
\brac{\Xi_{m} (x) \cdot \Xi (x) } 
\cdot 
\brac{\Xi_{m} (x) \cdot \Xi (x) }^\dagger
\nn\\
\in&
\brab{X_{(3)}^A , S^A}
}
is expanded in terms of 
$X_{(3)}^A$ and $S^A (=T_{\rm extra}^A)$  only because
the Maurer-Cartan 1-forms of $\Xi_{m} (x) \cdot \Xi (x)$ are constructed by the broken generators corresponding to $G_{\rm global}/H$ and $H_{\rm extra}$.
This relation yield  
\bee{
\hat{\alpha}_{\mu \perp (1)}(x)+\hat{\alpha}_{\mu \perp (1)}^{(m)} (x) & = 0 \ , 
\notag\\
\hat{\alpha}_{\mu \perp (2)}(x)+\hat{\alpha}_{\mu \perp (2)}^{(m)} (x) & =0 \ ,
\notag\\
\hat{\alpha}_{\mu \perp (1)}^{(I=0)}(x)+\hat{\alpha}_{\mu \perp (1)}^{(m)(I=0)} (x) & = 0 \ ,
\notag\\
\hat{\alpha}_{\mu \perp (2)}^{(I=0)}(x)+\hat{\alpha}_{\mu \perp (2)}^{(m)(I=0)} (x) & =0 \ .
}

To take account of the effect of  current quark masses, 
we introduce an external source $\chi$ which transforms under the chiral symmetry $G_{\rm global}$ as
\bee{
\chi \ra& \mc{G}\cdot \chi \cdot \mc{G}^\dagger \ .
}
We assume that its expectation value is given as
\bee{
\brae{\chi}=m_\pi^2 ~\bar{\Sigma}
=\braa{\mt{0 & m_\pi^2 1_2\\ m_\pi^2 1_2&0}}
\ .
}
We redefine the external source as
\bee{
\hat{\chi}
\equiv
\Xi_{m} (x)\cdot
\Xi (x)
\cdot\chi\cdot 
\Xi^\dagger (x)
\cdot
\Xi_{m}^\dagger (x)
}
such that it transforms 
under $G_{\rm global}\times G_{\rm local}$, $\mc{P}$, and $\mc{C}$ as
\bee{
\hat{\chi} \ra&~
\tilde{h} (x)\cdot 
\hat{\chi}\cdot 
\tilde{h}^\dagger (x)
\ ,\nn\\
\hat{\chi} \rax{\mc{P}}{}&~
\bar{\Sigma}\cdot 
\hat{\chi}\cdot \bar{\Sigma}
\ ,\nn\\
\hat{\chi} \rax{\mc{C}}{}&~
\bar{\Sigma}\cdot 
\braa{\hat{\chi}}^*\cdot \bar{\Sigma}
\ ,
}
respectively.

By using a standard order counting manner for the fields:
\bee{
V_\mu 
\sim 
\tilde{V}_\mu 
\sim 
\mc{O} (p)
\ ,~~
\alpha_\mu  
\sim 
\mc{O} (p)
\ ,~~
\hat{\chi}
\sim 
\mc{O} (p^2)
\ ,
}
possible operators invariant 
under $G_{\rm global} \times G_{\rm local} (\times H_{\rm extra} )$ as well as $\mc{P}$ and $\mc{C}$ at $\mc{O}(p^2)$ 
which do not include $\hat{\alpha}_{\parallel}^{(m)} (x)$ are $\hat{\alpha}_{\parallel}^{(m)(I=0)} (x)$ 
are written as
\bee{
\mc{L}_{1}=&
F^2\tr\brac{\braa{\hat{\alpha}_{\mu \perp (3)} (x)+\hat{\alpha}_{\mu \perp (3)}^{(m)} (x)}^2}
\ ,
\nn\\
\mc{L}_{2}=&
F^2\tr\brac{\braa{\hat{\alpha}_{\mu \parallel} (x)}^2}
\ ,
\nn\\
\mc{L}_{3}=&
F^2\tr\brac{\braa{\hat{\alpha}_{\mu \perp (1)}^{(m)} (x)}^2}
\ ,
\nn\\
\mc{L}_{4}=&
F^2\tr\brac{\braa{\hat{\alpha}_{\mu \perp (2)}^{(m)} (x)}^2}
\ ,
\nn\\
\mc{L}_{5}=&
F^2\tr\brac{\braa{\hat{\alpha}_{\mu \perp (3)}^{(m)} (x)}^2}
\ ,
\nn\\
\mc{L}_{6}=&
2F^2\tr\brac{\hat{\alpha}_{\mu \parallel}^{(m)} (x) \cdot  \hat{\alpha}^{{(m)} \mu}_{\perp (1)} (x)\cdot \bar{\Sigma}}
\ ,
\nn\\
\mc{L}_{7}=&
-2F^2\tr\brac{\braa{\hat{\alpha}_{\mu \perp (3)} (x)+\hat{\alpha}_{\mu \perp (3)}^{(m)} (x)} \cdot \hat{\alpha}^{{(m)} \mu}_{\perp (3)} (x)}
\ ,
\nn\\
\mc{L}_{8}=&
F^2\tr\brac{\braa{\hat{\alpha}_{\mu \perp (3)}^{(I=0)} (x)+\hat{\alpha}_{\mu \perp (3)}^{(m) (I=0)} (x)}^2}
\ ,
\nn\\
\mc{L}_{9}=&
F^2\tr\brac{\braa{\hat{\alpha}_{\mu \parallel}^{ (I=0)} (x)}^2}
\ ,
\nn\\
\mc{L}_{10}=&
F^2\tr\brac{\braa{\hat{\alpha}_{\mu \perp (1)}^{(m) (I=0)} (x)}^2}
\ ,
\nn\\
\mc{L}_{11}=&
F^2\tr\brac{\braa{\hat{\alpha}_{\mu \perp (2)}^{(m) (I=0)} (x)}^2}
\ ,
\nn\\
\mc{L}_{12}=&
F^2\tr\brac{\braa{\hat{\alpha}_{\mu \perp (3)}^{(m) (I=0)} (x)}^2}
\ ,
\nn\\
\mc{L}_{13}=&
2F^2\tr\brac{\hat{\alpha}_{\mu \parallel}^{ (I=0)} (x) \cdot  \hat{\alpha}^{{(m)} (I=0) \mu}_{\perp (1)} (x)\cdot \bar{\Sigma}}
\ ,
\nn\\
\mc{L}_{14}=&
-2F^2\tr \Big[
\braa{\hat{\alpha}_{\mu \perp (3)}^{(I=0)} (x)+\hat{\alpha}_{\mu \perp (3)}^{(m) (I=0)} (x)}
\notag\\
& \qquad\qquad\qquad
\cdot  \hat{\alpha}^{{(m)} (I=0) \mu}_{\perp (3)} (x) \Big]
\ ,
\nn\\
\mc{L}_{\chi}=&
F^2\tr\brac{\hat{\chi}\cdot \bar{\Sigma}}
\label{Lagrangian operators1}
\ ,
}
where $F$ is a constant with dimension one.  Note that this $F$ is not the pion decay constant, which will be determined later.
In addition, 
there are eight operators including $\hat{\alpha}_{\parallel}^{(m)} (x)$ or $\hat{\alpha}_{\parallel}^{{(m)}(I=0)} (x)$:
\bee{
{\mc{L}}'_{1}=&
F^2\tr\brac{\braa{\hat{\alpha}_{\mu \parallel}^{(m)} (x)}^2}
\ ,
\nn\\
{\mc{L}}'_{2}=&
2F^2\tr\brac{\hat{\alpha}_{\mu \parallel}^{(m)} (x) \cdot \hat{\alpha}_{ \parallel}^{\mu} (x)}
\ ,
\nn\\
{\mc{L}}'_{3}=&
2F^2\tr\brac{\hat{\alpha}_{\mu \parallel}^{(m)} (x) \cdot  \hat{\alpha}^{ \mu}_{\perp (1)} (x)\cdot \bar{\Sigma}}
\ ,
\nn\\
{\mc{L}}'_{4}=&
2F^2\tr\brac{\hat{\alpha}_{\mu \parallel}^{(m)} (x) \cdot  \hat{\alpha}^{{(m)} \mu}_{\perp (1)} (x)\cdot \bar{\Sigma}}
\ ,
\nn\\
{\mc{L}}'_{5}=&
F^2\tr\brac{\braa{\hat{\alpha}_{\mu \parallel}^{(m) (I=0)} (x)}^2}
\ ,
\nn\\
{\mc{L}}'_{6}=&
2F^2\tr\brac{\hat{\alpha}_{\mu \parallel}^{(m) (I=0)} (x) \cdot \hat{\alpha}_{ \parallel}^{(I=0)\mu} (x)}
\ ,
\nn\\
{\mc{L}}'_{7}=&
2F^2\tr\brac{\hat{\alpha}_{\mu \parallel}^{(m) (I=0)} (x) \cdot  \hat{\alpha}^{ (I=0) \mu}_{\perp (1)} (x)\cdot \bar{\Sigma}}
\ ,
\nn\\
{\mc{L}}'_{8}=&
2F^2\tr\brac{\hat{\alpha}_{\mu \parallel}^{(m) (I=0)} (x) \cdot  \hat{\alpha}^{{(m)} (I=0) \mu}_{\perp (1)} (x)\cdot \bar{\Sigma}}
\label{Lagrangian operators3}
\ .
}
Although the term given by
\bee{
F^2\tr\brac{\hat{\chi}}
}
is also an allowed operator, it contributes only to the vacuum energy.
The Lagrangian with the $\mbox{SU(4)}\times \mbox{U(1)}$ HLS at $\mc{O}(p^2)$ is written as
\bee{
\tilde{\mc{L}}_{\brac{\rm SU(4)\times U(1)}_{\rm HLS}}^{\mc{O} (p^2) }=&
\sum_{n=1}^{14} {a}_{(n)} {\mc{L}}_n
+
\sum_{n=1}^{8} {b}_{(n)} {{\mc{L}}'}_n
\nn\\
&
-
\frac{1}{2g^2}\tr\brac{V_{\mu\nu}V^{\mu\nu}}
\nn\\
&
-
\braa{\frac{1}{g^2_B} -\frac{1}{g^2}}\tr\brac{V_{\mu\nu}}\tr\brac{V^{\mu\nu}}
\nn\\
&
+
a_\chi\mc{L}_{\chi}
\label{Lagrangian full}
}
with arbitrary real coefficients ${a}_{(n)}$, ${b}_{(n)}$, and $a_\chi$.
$g$ and $g_B$ are the gauge coupling corresponding to 
$\mbox{SU(4)}_{\rm HLS}$ and $\mbox{U(1)}_{\rm HLS}$, respectively\footnote{The values of the couplings are scaled by $\sqrt{2}$ comparing with usual way as defined in Ref.\,\cite{Harada:2003jx}.}.

In the above Lagrangian (\ref{Lagrangian full}), 
$\Xi (x)$ and $\Xi_m (x)$ are parametrized as
\bee{
\Xi (x)\equiv &
e^{ -i p /F_p} \cdot e^{ i s /F_s } \cdot  e^{i \pi /F_\pi }  
= \Xi \braa{p}^\dagger \cdot\Xi \braa{s} \cdot \Xi \braa{\pi}
\ ,
\nn\\
\Xi_{m} (x)\equiv &e^{ -i \tilde{s} /F_{\tilde{s}}}\cdot e^{ i p /F_p}  
=\Xi^\dagger \braa{\tilde{s}} \cdot  \Xi \braa{p}
\label{parametrizing}
}
where the $p$, $s$, $\tilde{s}$, and $\pi$ are
\bee{
p (x) =&  
p^A_{(1)} (x) \cdot X_{(1)}^A
+ p^A_{(2)} (x) \cdot X_{(2)}^A 
+p^A_{(3)} (x)\cdot X_{(3)}^A
\ ,
\nn\\
s (x) =& s^A_s (x) \cdot S^A 
\ ,~~~~
\tilde{s} (x) = \tilde{s}^A_s (x) \cdot S^A 
\ ,~~~~
\nn\\
\pi (x)=&\pi^A (x)\cdot X^A_{(3)}
\ ,
}
respectively.
$\pi (x)$ is the NG-boson field corresponding to the breaking of the chiral symmetry, which is 
identified with the  pion.
$F_p$, $F_s$, $F_\pi$, and $F_{\tilde{s}}$ are constants with one mass-dimension, in particular $F_\pi$ is the pion decay constant.
$p(x)$, $s(x)$, and $\tilde{s}(x)$ are also the NG-boson fields which are eaten by the gauge fields.

Since $\tilde{V}_\mu$ included in $\hat{\alpha}_{\mu \parallel}^{(m)}$ is 
not a dynamical field, we fix $\tilde{s}(x)=0$ and integrate out the gauge field.
Then $\mc{L}'_n$ become the terms given in Eqs.~\eqref{Lagrangian operators1}, and the Lagrangian is written by 
\begin{equation}
{\mc{L}}_{\brac{\rm SU(4)\times U(1)}_{\rm HLS}}^{\mc{O} (p^2) }= {\mathcal L}_V + {\mathcal L}_k + a_\chi{\mathcal L}_{\chi} \ , 
\label{Lagrangian full2}
\end{equation}
where
\begin{align}
{\mathcal L}_V = & \sum_{n=1}^{14} \bar{a}_{(n)} {\mc{L}}_n
\nn\\
{\mathcal L}_k = & 
- \frac{1}{2g^2}\tr\brac{V_{\mu\nu}V^{\mu\nu}}
\notag\\
& - \braa{\frac{1}{g^2_B} -\frac{1}{g^2}}\tr\brac{V_{\mu\nu}}\tr\brac{V^{\mu\nu}}
\ ,
\label{Lagrangian terms}
\end{align}
with the coefficients $\bar{a}_{(n)}$ being certain 
linear combinations of $a_{(n)}$ and $b_{(n)}$.

To analyze the dynamics of the spin-1 mesons together with the pion in the model, 
in the following analysis, we take the unitary gauge
\bee{
p=s=0
}
as well as $\tilde{s} =0$.
As shown in  Appendix~\ref{sec:expansion},  the expanded form of the Maurer-Cartan 1-forms are written by using $\pi$ and $V_\mu$.

\section{Eigenstates and masses}
\label{sec:Vector mesons and pion}

In this section, 
we obtain the mass eigenstates of the spin-1 mesons and their masses.

By using the generators of $\mbox{SU(4)}\times \mbox{U(1)}$ listed in Appendix~\ref{sec:Generators}, the 
HLS gauge field is decomposed as
\bee{
V_{\mu}=&
V_{\mu \parallel}
+ V_{\mu \perp (1)}
+ V_{\mu \perp (2)}
+ V_{\mu \perp (3)}
\nn\\
&
+
V_{\mu \parallel}^{(I=0)}
+
V_{\mu \perp (1)}^{(I=0)}
+
V_{\mu \perp (2)}^{(I=0)}
+
V_{\mu \perp (3)}^{(I=0)}
\ ,
}
where 
\bee{
V_{\mu \parallel}
\equiv&
2\tr\brac{V_{\mu} \cdot S^a}S^a
\equiv
V_{\mu \parallel}^a \cdot S^a
\ ,
\nn\\
V_{\mu \parallel}^{(I=0)}
\equiv&
2\tr\brac{V_{\mu} \cdot S^0}S^0
\equiv
V_{\mu \parallel}^0 \cdot S^0
\ ,
\nn\\
V_{\mu \perp (i)}
\equiv&
2\tr\brac{V_{\mu} \cdot X^a_{(i)}}X^a_{(i)}
\equiv
V_{\mu \perp}^a \cdot X^a_{(i)}
\ ,
\nn\\
V_{\mu \perp (i)}^{(I=0)}
\equiv&
2\tr\brac{V_{\mu} \cdot X^0_{(i)}}X^0_{(i)}
\equiv
V_{\mu \perp}^0 \cdot X^0_{(i)}
\ .
}
They are classified as $\braa{\rho, \omega, a_1, \rho', \omega', b_1, f_1, h_1}$
by the properties of transformation under $\mc{P}$ and $\mc{C}$.
Because the fields satisfy 
\bee{
V_\mu \rax{\mc{P}}{}&
\bar{\Sigma}\cdot V^{\mu}\cdot \bar{\Sigma} 
\nn\\
=&
V^{\mu}_{\parallel}
+
V^\mu_{\perp (1)}
-
V^\mu_{\perp (2)}
-
V^\mu_{\perp (3)}
\nn\\
&+
V_{\parallel}^{\mu (I=0)}
+
V_{\perp (1)}^{\mu (I=0)}
-
V_{\perp (2)}^{\mu (I=0)}
-
V_{\perp (3)}^{\mu (I=0)}
\ ,
\nn\\
V_\mu \rax{\mc{C}}{}&
-
\bar{\Sigma}\cdot \braa{V_{\mu}}^*\cdot \bar{\Sigma} 
\nn\\
=&
-
\braa{V_{\mu \parallel}}^*
-
\braa{V_{\mu \perp (1)}}^*
-
\braa{V_{\mu \perp (2)}}^*
+
\braa{V_{\mu \perp (3)}}^*
\nn\\
&
-
\braa{V_{\mu \parallel}^{(I=0)}}^*
-
\braa{V_{\mu \perp (1)}^{(I=0)}}^*
-
\braa{V_{\mu \perp (2)}^{(I=0)}}^*
+
\braa{V_{\mu \perp (3)}^{(I=0)}}^*
\ ,
}
they are identified as the spin-1 mesons:
\bee{
&
\braa{V_{\mu \parallel} \ , V_{\mu \perp {(1)}}}
\Ra \rho
\ , 
\rho'
\ , ~~
\notag\\
&
\braa{V_{\mu \parallel}^{(I=0)} \ , V_{\mu \perp {(1)}}^{(I=0)}}
\Ra \omega
\ , 
\omega'
\ ,
\nn\\
&
V_{\mu \perp {(2)}} \Ra b_1
\ , ~~
V_{\mu \perp {(2)}}^{(I=0)} \Ra h_1
\ , ~~
\notag\\
&
V_{\mu \perp {(3)}} \Ra a_1
\ , ~~
V_{\mu \perp {(3)}}^{(I=0)} \Ra f_1
\label{def of the vector meson}
\ .
}
\begin{widetext}
From ${\mc{L}}_{V}+ a_\chi{\mathcal L}_{\chi} $, the quadratic terms with respect to  the fields are given as
\bee{
 {\mc{L}}_{V}+ a_\chi{\mathcal L}_{\chi} 
=&
\frac{1}{2}\frac{F^2}{F_\pi^2}
\braa{\bar{a}_{(1)}-\frac{\bar{a}_{(7)}^2}{\bar{a}_{(5)} }}\braa{\pd_\mu \pi^a}^2
+
\frac{1}{2}\braa{a_{\chi}\frac{F^2}{F_\pi^2}} m_\pi^2 \braa{\pi^a}^2
+
\frac{1}{2}\frac{F^2}{F_\pi^2}
\braa{\bar{a}_{(8)}-\frac{\bar{a}_{(14)}^2}{\bar{a}_{(12)} }} \braa{\pd_\mu \eta}^2
+
\frac{1}{2}\braa{a_{\chi}\frac{F^2}{F_\pi^2}} m_\pi^2 \braa{\eta}^2
\nn\\
&+
\sqrt{2}\frac{F^2}{F_\pi}
\braa{\bar{a}_{(1)}-\frac{\bar{a}_{(7)}^2}{\bar{a}_{(5)} }}
\braa{ \mc{A}_\mu^a \pd^\mu \pi^a}
+
\sqrt{2}\frac{F^2}{F_\pi}
\braa{\bar{a}_{(8)}-\frac{\bar{a}_{(14)}^2}{\bar{a}_{(12)} }}
\braa{ \mc{A}_\mu^0 \pd^\mu \eta}
\nn\\
&+
\frac{1}{2}
\braa{\bar{a}_{(4)}F^2g^2}
\frac{1}{g^2}\braa{V_{\mu \perp{(2)}}^a }^2
+
\frac{1}{2}
\braa{\bar{a}_{(11)}F^2g^2}
\frac{1}{g^2}\braa{V_{\mu \perp{(2)}}^0}^2
\nn\\
&
+
\frac{1}{2}
\braa{\bar{a}_{(5)}F^2g^2}
\frac{1}{g^2}\braa{V_{\mu \perp{(3)}}^a -\frac{\bar{a}_{(7)}}{\bar{a}_{(5)}} \frac{1}{ F_\pi}\pd_\mu \pi^a}^2
+
\frac{1}{2}
\braa{\bar{a}_{(12)}F^2g^2}
\frac{1}{g^2}\braa{V_{\mu \perp{(3)}}^0 - \frac{\bar{a}_{(14)}}{\bar{a}_{(12)}}\frac{1}{ F_\pi}\pd_\mu \eta}^2
\nn\\
&
+
g^2F^2
\tr\brac{
\braa{\mt{
\frac{1}{g}V_{\mu \parallel} &
\frac{1}{g}\bar{\Sigma}\cdot V_{\mu \perp{(1)}}
}}
\braa{\mt{
\bar{a}_{(2)}&-\bar{a}_{(6)}
\\
-\bar{a}_{(6)}&\bar{a}_{(3)}
}}
\braa{\mt{
\frac{1}{g}V_{\mu \parallel} \\
\frac{1}{g}\bar{\Sigma}\cdot V_{\mu \perp{(1)}}
}}
}
\nn\\
&
+
g^2F^2
\tr\brac{
\braa{\mt{
\frac{1}{g_B}V_{\mu \parallel}^{(I=0)} &
\frac{1}{g}\bar{\Sigma}\cdot V_{\mu \perp{(1)}}^{(I=0)}
}}
\braa{\mt{
\frac{g_B^2}{g^2}\bar{a}_{(9)}&-\frac{g_B}{g}\bar{a}_{(13)}
\\
-\frac{g_B}{g}\bar{a}_{(13)}&\bar{a}_{(10)}
}}
\braa{\mt{
\frac{1}{g_B}V_{\mu \parallel}^{(I=0)} \\
\frac{1}{g}\bar{\Sigma}\cdot V_{\mu \perp{(1)}}^{(I=0)}
}}
}
+
\cdots
\label{L_V and L_chi}
}
where the axial external gauge 
field is defined as 
$\mc{A}^a_\mu =\frac{1}{2}\braa{\mc{R}_\mu^a - \mc{L}_\mu^a}$. 
Note that the field $\eta$ defined by $\eta \equiv 2\tr\brac{\pi \cdot X^0_{(3)}}$ is the linear combination of the lowest eta meson and $\eta'(958)$.
To normalize the kinetic terms of $\pi^a$ and $\eta$, we set 
\bee{
F^2\braa{\bar{a}_{(1)}-\frac{\bar{a}_{(7)}^2}{\bar{a}_{(5)} }}=&F_\pi^2
\ ,~~~
F^2\braa{\bar{a}_{(8)}-\frac{\bar{a}_{(14)}^2}{\bar{a}_{(12)} }}=F_\pi^2
\ ,
\label{normalization of pion}
}
together with $a_{\chi}F^2 =F_\pi^2$ which makes the pion mass be  $m_\pi$.  
The second line in Eq.\,\eqref{L_V and L_chi} implies that the physical pion decay constant is defined as 
\bee{
f_\pi \equiv \sqrt{2}F_\pi
\ ,
}
whose value is given in Table\,\ref{Table:Experimental values}.
Furthermore, the mass eigenstates of $a_1$ and $f_1$ are defined by 
\bee{
\braa{a_1}_\mu \equiv& \frac{1}{g} \braa{V_{\mu \perp{(3)}}^a - \frac{r_{a_1}}{ f_\pi}\pd_\mu \pi^a}X_{(3)}^a
\ ,
~~
\braa{f_1}_\mu \equiv \frac{1}{g} \braa{V_{\mu \perp{(3)}}^0 - \frac{r_{f_1}}{ f_\pi}\pd_\mu \eta}X_{(3)}^0
\label{definition: a1 f1}
\ ,
}
where 
\bee{
r_{a_1}\equiv&
\sqrt{2}\frac{\bar{a}_{(7)}}{\bar{a}_{(5)}}
\ ,~~~~
r_{f_1}\equiv
\sqrt{2}\frac{\bar{a}_{(14)}}{\bar{a}_{(12)}}
\ .
}
$r_{a_1}$ ($r_{f_1}$) expresses the mixing rate between the $a_1$ ($f_1$) meson and the pion $\pi$ ($\eta$).
Their masses are obtained as
\bee{
m_{a_1}^2 = 
\bar{a}_{(5)}g^2F^2
\ ,~~~
m_{f_1}^2 = 
\bar{a}_{(12)}g^2F^2
\ .
\label{masses af}
}
The mixing of Eq.\,\eqref{definition: a1 f1}
implies that the currents corresponding to the generators  $X_{(3)}^a$ and $X_{(3)}^0$  of the \mbox{SU(4)} HLS are coupled to the axial current of the chiral symmetry with the factors of $r_{a_1}$ and $r_{f_1}$, respectively.   
\end{widetext}

The physical states and masses for $b_1$ and $h_1$ are defined as
\bee{
\braa{b_1}_\mu \equiv& \frac{1}{g}V_{\mu \perp {(2)}} 
~~~~
m_{b_1}^2=\bar{a}_{(4)}g^2F^2
\ , \nn\\
\braa{h_1}_\mu \equiv& \frac{1}{g}V_{\mu \perp {(2)}}^{(I=0)} 
~~~~
m_{h_1}^2=\bar{a}_{(11)}g^2F^2
\ ,
\label{masses bh}
}
respectively.

By diagonalizing the mass matrices from Eq.\,\eqref{L_V and L_chi}, the eigenstates for the vector mesons are expressed as
\begin{align}
\rho_\mu &= \rho_\mu^a S^a \equiv \frac{1}{g}\braa{\cos \theta_\rho V_{\mu \parallel} -\sin \theta_\rho \bar{\Sigma}\cdot V_{\mu \perp{(1)}}}
\ , \notag\\
\braa{\rho'}_\mu &= \braa{\rho'}_\mu^a S^a\equiv \frac{1}{g}\braa{\cos \theta_\rho \bar{\Sigma}\cdot V_{\mu \perp{(1)}} + \sin \theta_\rho V_{\mu \parallel}}
\ , \notag\\
\hat{\omega}_\mu &= \omega_\mu S^0 \equiv \frac{1}{g_B}\cos \theta_\omega V_{\mu \parallel}^{(I=0)} -\frac{1}{g}\sin \theta_\omega \bar{\Sigma}\cdot V_{\mu \perp{(1)}}^{(I=0)}
\ , \notag\\
\hat{\braa{\omega'}}_\mu &= \braa{\omega'}_\mu S^0 \equiv \frac{1}{g}\cos \theta_\omega \bar{\Sigma}\cdot V_{\mu \perp{(1)}}^{(I=0)} + \frac{1}{g_B}\sin \theta_\omega V_{\mu \parallel}^{(I=0)}
\ .
\end{align}
The masses of these states are obtained as
\bee{
m_{\rho , \rho'}^2
\equiv&
\frac{1}{2}\braa{\bar{a}_{(2)}+\bar{a}_{(3)}
\mp \sqrt{\braa{\bar{a}_{(2)} - \bar{a}_{(3)}}^2+4\bar{a}_{(6)}^2}}g^2F^2
\ ,\nn\\
m_{\omega , \omega'}^2
\equiv&
\frac{1}{2}\braa{\frac{g_B^2}{g^2}\bar{a}_{(9)}+\bar{a}_{(10)}
\right.
\nn\\&
\left.\mp\sqrt{\braa{\frac{g_B^2}{g^2}\bar{a}_{(9)}-\bar{a}_{(10)}}^2+4\frac{g_B^2}{g^2}\bar{a}_{(13)}^2}}g^2F^2
\ ,
\label{masses rho omega}
}
where
 the mixing angles $\theta_\rho$ and $\theta_\omega$ are determined as 
\bee{
\tan 2\theta_\rho \equiv& 
\frac{2\bar{a}_{(6)} }{\bar{a}_{(2)} - \bar{a}_{(3)}}
\ ,~~~
\tan 2\theta_\omega \equiv
\frac{2g_B g\bar{a}_{(13)} }{g^2_B\bar{a}_{(9)}-g^2\bar{a}_{(10)}}
\ .
}

We can fix the values of ten parameters 
from the physical values of eight spin-1 mesons listed in Table~\ref{Table:masses} 
 using Eqs.~(\ref{masses af}), (\ref{masses bh}) and (\ref{masses rho omega}) together with 
two conditions given in Eq.~\eqref{normalization of pion}.
The model still has six free parameters:
\bee{
g
\ ,~~
g_B
\ ,~~
r_{a_1}
\ ,~~
r_{f_1}
\ ,~~
\cos \theta_\rho
\ ,~~
\cos \theta_\omega
\ ,
}
which relates several interactions among the spin-1 mesons.
In this paper, since we do not treat decays of the eta meson, 
the parameter $r_{f_1}$ is irrelevant.
So, 
we will determine the values of five parameters except for $r_{f_1}$.
\begin{table}
\caption[]{Masses of the  relevant  spin-1 mesons in PDG~\cite{Agashe:2014kda}}
\begin{center}
\begin{tabular}{c|cc}\hline\hline
Mesons &$I\braa{J^{PC}}$&mass (MeV)
\\\hline
$\rho$&$1\braa{1^{--}}$&$775.26\pm0.25$
\\
$\omega$&$0\braa{1^{--}}$&$782.65\pm0.12$
\\
$\rho'$&$1\braa{1^{--}}$&$1465\pm 25$
\\
$\omega'$&$0\braa{1^{--}}$&$1400$ -- $1450$
\\
$a_1$&$1\braa{1^{++}}$&$1230\pm 40$
\\
$f_1$&$0\braa{1^{++}}$&$1281.9 \pm 0.5$
\\
$b_1$&$1\braa{1^{+-}}$&$1229.5\pm3.2$
\\
$h_1$&$0\braa{1^{+-}}$&$1170\pm20$
\\\hline\hline
\end{tabular}
\end{center}
\label{Table:masses}
\end{table}
\begin{table}
\caption[]{Experimental values from PDG~\cite{Agashe:2014kda}.}
\begin{center}
\begin{tabular}{c|c}\hline\hline
$\Gamma \braa{\rho \ra \pi\pi}
$&$
147.8 \pm 0.9 \,{\rm MeV}
$
\\
$\Gamma \braa{\rho^0 \ra e^+e^-}
$&$
7.04 \pm 0.06 \,{\rm keV}
$
\\
$\Gamma(\omega \to e^+ e^-)$ & $0.60 \pm 0.02\,\mbox{keV}$
\\
$m_{\pi^\pm}
$&$
139.57018 \pm 0.00035 \,{\rm MeV}
$\\
$m_e
$&$
548.57990946 \pm 0.00000022 \,{\rm keV}
$\\
$\alpha\equiv\frac{e^2}{4\pi}
$&$
\frac{1}{137}
$\\
$f_\pi
$&$
92.21 \pm 0.14\,{\rm MeV}
$
\\
$\brae{r^2}^{\pi^\pm}_V
$&$
0.452 \pm 0.011
\,{\rm fm^2}
$
\\\hline\hline
\end{tabular}
\end{center}
\label{Table:Experimental values}
\end{table}

In the following, we summarize the extended GT relations, the 
relations among one-pion decays of spin-1 mesons
and the extended KSRF relations in the separated sections.
To obtain some predictions analytically and numerically from them, we use experimental values in Table\,\ref{Table:Experimental values}.

\section{Extended Goldberger-Treiman relation} 
\label{sec:GT}

In this section, we investigate an  extended Goldberger-Treiman relation for one-pion interactions of two different spin-1 mesons.
First, we give a general discussion for the extended GT relation. 
Next, we  derive several relations  in the HLS model. 

Let us start a general 
discussion in the case  that the final spin-1 meson state has the different parity from   the initial state.
By requiring 
Lorentz covariance and parity invariance, the
amplitudes of two spin-1 states coupled with the axial current $j_5^\alpha$ 
are written as\,\footnote{%
Indices for isospin are omitted. Note that terms including 
the antisymmetric tenser $\epsilon^{\mu\nu\alpha\beta }$ are also allowed if
the initial and final states have the same parity.
In the HLS model, 
these contributions are obtained from the intrinsic parity odd terms,
which are  listed  in Appendix\,\ref{sec:Intrinsic parity odd terms}.
}
\bee{
\mc{M}^\alpha 
=&
\int d^4x e^{-iqx}\bra{V_\mu (p_2)} j^\alpha_5 (x) \ket{V_\nu (p_1)}
\nn\\
=&
\epsilon^*_\mu (p_2)
\brac{
g_1 (q^2 )
g^{\mu\nu} 
i p^\alpha 
+
g_2(q^2 ) 
\frac{q^{\mu}q^{\nu}}{m^2_1 + m_2^2}
i p^\alpha 
\right.\nn\\&
~~~~~~~
+
g_3(q^2 ) 
 \braa{
iq^{\mu}g^{\nu \alpha} 
+
iq^{\nu}g^{\mu \alpha} 
}
\nn\\&
~~~~~~~
+
g_4(q^2 ) 
 \braa{
 i q^{\mu}g^{\nu \alpha} 
-
i q^{\nu}g^{\mu \alpha} 
}
\nn\\&\left.
~~~~~~
+
h_1 (q^2 )
 g^{\mu\nu}i q^\alpha
+
h_2 (q^2 )
 \frac{q^{\mu}q^{\nu}}{m^2_1 + m_2^2} i q^\alpha 
}
\epsilon_\nu (p_1)
\label{axial form factor}
}
where $p= \frac{p_1+p_2}{2}$, $q= p_1-p_2$, and $g_i(q^2)$ ($i=1,2,3,4$) and $h_j(q^2)$ ($j=1,2$) are 
 independent form factors, which are generally complex functions of $q^2$. 
Since the axial vector current is conserved in the chiral limit, we have the Ward-Takahashi identity as   
$q_\alpha \mc{M}^\alpha = 0 $,
which leads to
\bee{
&
g_1 (q^2 ) \braa{p\cdot q}
\braa{\epsilon^*  \cdot \epsilon}
+
 g_2 (q^2 ) 
\braa{p\cdot q}
\frac{\braa{\epsilon^* \cdot q}
\braa{q \cdot \epsilon}}{m_1^2+m_2^2}
\nn\\&
+
2g_3 (q^2 )
\braa{\epsilon^* \cdot q}
\braa{q \cdot \epsilon}
\nn\\&
+
 h_1 (q^2 )  q^2
\braa{\epsilon^*  \cdot \epsilon}
+
 h_2 (q^2 ) q^2
\frac{\braa{\epsilon^* \cdot q}
\braa{q \cdot \epsilon}}{m_1^2+m_2^2}
=0
\ .
\label{WT identity}
}
The form factors 
$h_1 (q^2)$ and $h_2 (q^2)$ include a massless pole of the pion contribution:
\bee{
h_{n} (q^2) = \frac{f_\pi}{q^2} G_{V_1V_2\pi}^{(n)} + \cdots
\label{hn}
}
with $n=1,2$.   
In the soft pion limit $q^2 \ra 0$, 
the left hand side of Eq.~(\ref{WT identity}) is reduced to
\begin{align}
&
\mbox{(LHS of Eq.~(\ref{WT identity}))} \notag\\
& \quad = \bigg[ g_1 (0) \frac{m_1^2-m_2^2}{2} 
 + f_\pi \, G_{V_1V_2 \pi}^{(1)} \bigg] \left( \epsilon^\ast \cdot \epsilon \right)  \ ,
\end{align}
where we used 
$p\cdot q=\frac{m_1^2 -m_2^2}{2}$. From this
one can obtain 
\bee{
G_{V_1 V_2 \pi}^{(1)}
=&
-
\frac{m_1^2- m_2^2}{2f_\pi}
g_1 (0 )
\ .
\label{Extended GT relation from general discussion}
}
This is an extended Goldberger-Treiman relation among a mass difference of two spin-1 mesons, their coupling to one pion and 
the axial form factor.

It should be noted that, if 
the mass splitting of the initial and final states were large, the soft pion limit would   not be reasonable.
The existence of the emergent symmetry in QCD implies that
the mass difference of the spin-1 mesons $\braa{\rho, a_1, \rho', \omega', b_1, f_1, h_1}$ comes from the breaking of the chiral symmetry.
Thus, 
the emergent symmetry together with the chiral symmetry ensures  low energy theorems
for the  members of a multiplet of the symmetry.

Next, we turn to make an analysis based on the present model. 
In the $\mc{L}_V$ part of the Lagrangian\,(\ref{Lagrangian full2}), there are no interactions among two HLS gauge fields and one pion field.  However, due to the existence of the 
$a_1$-$\pi$ and $f_1$-$\eta$ mixings as shown in Eq.\,\eqref{definition: a1 f1}, the HLS gauge field $V_\mu$ include the fields for 
the physical pion  in addition to the physical
spin-1 mesons. 
Then, 
the interactions among two spin-1 mesons and one pion are generated from 
\bee{
\mc{L}_{\rm int}^{(3)}
=&
-\frac{1}{ig^2}
\tr\brac{\braa{\pd_\mu V_\nu - \pd_\nu V_\mu} \brac{V^\mu, V^\nu}}
}
included in $\mc{L}_k$ of Eq.\,\eqref{Lagrangian terms}. 
As a result, all the interactions among two spin-1 mesons and one pion are proportional to the ratio $r_{a_1}/f_\pi$.
The explicit forms of the effective vertices are written as
\bee{
&\Gamma^{\mu\nu} \brac{\braa{V_1}_\mu^a (p_1), \braa{V_2}_\nu^b (p_2) , \pi^c }
\nn\\
=&
g_{V_1 V_2 \pi}\,
\epsilon^{abc}
\braa{p_1^2 P^{\mu\nu}(p_1)-p_2^2 P^{\mu\nu}(p_2)}
\ ,\nn\\
&\Gamma^{\mu\nu} \brac{\braa{V_1}_\mu^a (p_1), \braa{V_2^{(I=0)}}_\nu (p_2) , \pi^b }
\nn\\
=&
g_{V_1 V_2 \pi}\,
\delta^{ab}
\braa{p_1^2 P^{\mu\nu}(p_1)-p_2^2 P^{\mu\nu}(p_2)}
\ ,
\label{effective vertex for VVpi}
}
where the projection operator is defined as
\bee{
P^{\mu\nu} (p) \equiv g^{\mu\nu} -\frac{p^\mu p^\nu}{p^2}
\ ,
}
and $g_{V_1 V_2 \pi}$ expresses the corresponding coupling:
\bee{
g_{\rho a_1 \pi}
=&
g_{\rho' h_1 \pi}
=
\frac{r_{a_1}}{\sqrt{2}f_\pi}\cos \theta_\rho
\ ,\nn\\
g_{\rho' a_1 \pi}
=&
-g_{\rho h_1 \pi}
=
\frac{r_{a_1}}{\sqrt{2}f_\pi}\sin \theta_\rho
\ ,\nn\\
g_{b_1 \omega \pi}
=&
-\frac{r_{a_1}}{\sqrt{2}f_\pi}\sin \theta_\omega
\ ,~~
g_{b_1 \omega' \pi}
=
\frac{r_{a_1}}{\sqrt{2}f_\pi}\cos \theta_\omega
\ .
\label{GT relation}
}

\begin{figure*}
(a)
\includegraphics[width=40mm]{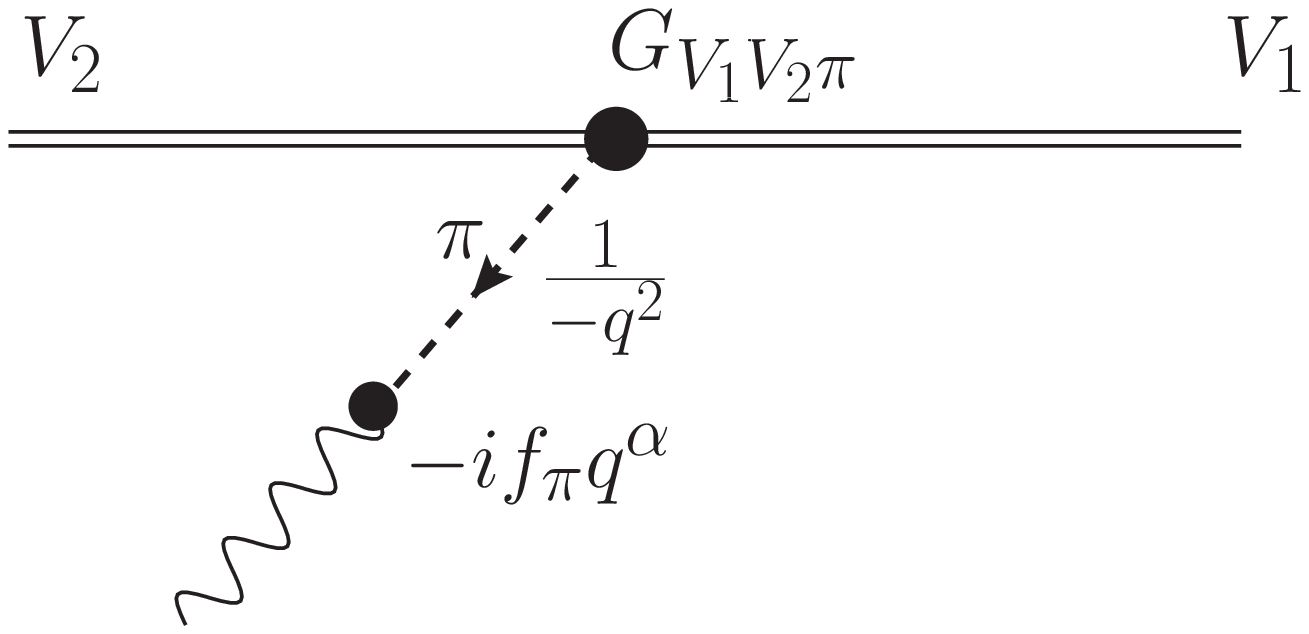}
~~(b)
\includegraphics[width=40mm]{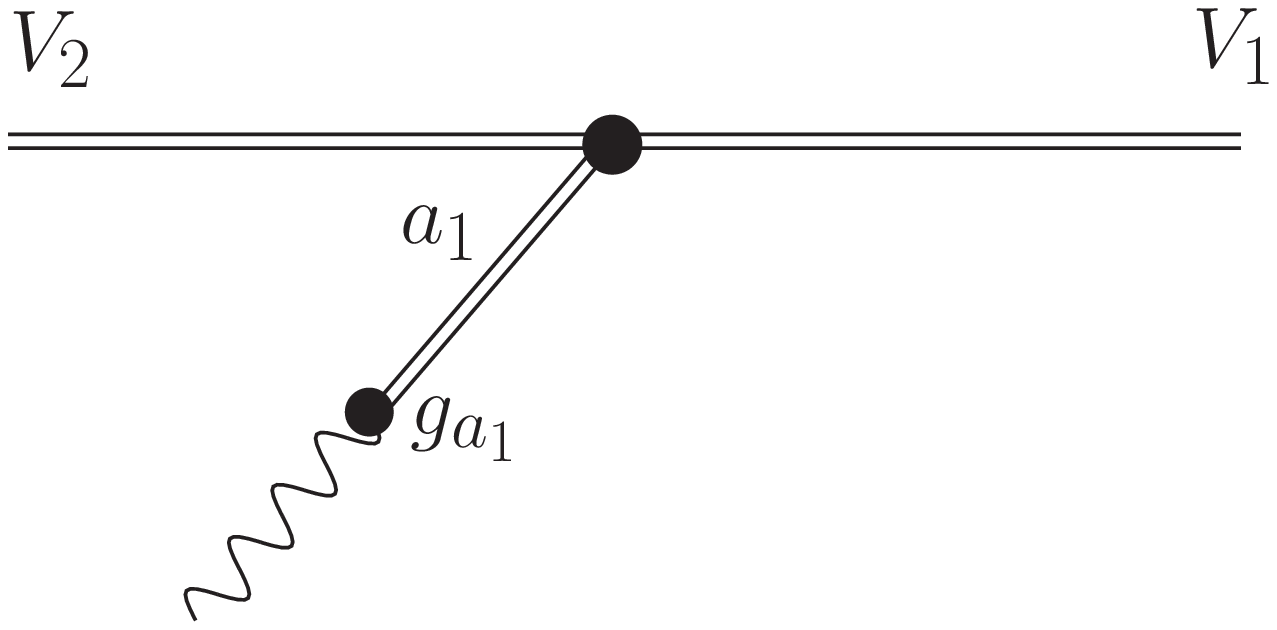}
\caption[]{Diagrams contributing to the amplitude given in Eq.\,\eqref{axial form factor}.}
\label{figure: couple to axial current}
\end{figure*}

As shown in Appendix\,\ref{sec:expansion}, interactions among three spin-1 mesons including $a_1$
are also obtained from $\mc{L}_{\rm int}^{(3)}$.
The  direct coupling  of two spin-1 mesons  with the axial external gauge field does not exist at the leading order of the present model, and only  two diagrams shown in FIG.\,\ref{figure: couple to axial current} 
contribute to the coupling to the axial vector current.  The pion in Fig.~\ref{figure: couple to axial current}(a) contributes to only $h_1$, and the $a_1$ meson in Fig.~\ref{figure: couple to axial current}(b) contributes to $h_1$, $g_1$ and $g_4$. We summarize their contributions in Table~\ref{Table:axial form factor}.  Substituting these contributions into Eq.~(\ref{WT identity}) we can easily verify that the Ward-Takahashi identity is actually satisfied for any $q^2$.

We next 
consider the soft-pion limit, $q^2 \to 0$. As expected in the general consideration given above, the pion contribution dominates over the $a_1$ meson contribution in $h_1$.  As a result, $h_1(q^2)$ is expressed as in Eq.~(\ref{hn}), where $G_{V_1V_2\pi}^{(1)}$ is listed in the first column of Table~\ref{Table:coupling VVpi}.  On the other hand, $g_1(0)$ is determined by taking $q^2 = 0$ limit of the $a_1$ meson contribution, which is listed in the second column of Table~\ref{Table:coupling VVpi}.
Since the coupling $g_{a_1}$ in the second column is given by $g_{a_1}= -\frac{r_{a_1} m_{a_1}^2}{g}$ as shown in Appendix\,\ref{sec:expansion},
we can easily confirm that these actually satisfy the extended GT relation in Eq.~\eqref{Extended GT relation from general discussion}.

\begin{table*}
\caption[]{Axial form factors given in the SU(4) HLS model. The function $D^{a_1} (q^2)$
is defined as $ \displaystyle D^{a_1} (q^2) \equiv \frac{m_{a_1}^2}{m_{a_1}^2-q^2}$.
We also find that the other form factors equal to zero at the $\mc{O}(p^2)$ order: $h_2(q^2)=g_2(q^2)=g_3(q^2)=0$. 
}
\begin{center}
\begin{tabular}{c|c|c|c}\hline\hline
&$h_1 (q^2)$&$g_1 (q^2)$&$g_4 (q^2)$
\\\hline
$a_1 \ra \rho$&$ \displaystyle
\braa{m_{a_1}^2-m_{\rho}^2}\braa{\frac{r_{a_1}}{\sqrt{2}f_\pi}\cos \theta_\rho}\frac{ f_\pi}{q^2}
-\frac{g}{\sqrt{2}}\cos\theta_\rho \frac{g_{a_1}}{m_{a_1}^2}
\frac{m_{a_1}^2-m_{\rho}^2}{m_{a_1}^2}
D^{a_1} (q^2)
$&$ \displaystyle
\sqrt{2} g\cos\theta_\rho 
\frac{g_{a_1}}{m_{a_1}^2}
D^{a_1} (q^2)
$&$ \displaystyle
\sqrt{2} g\cos\theta_\rho \frac{g_{a_1}}{m_{a_1}^2}
D^{a_1} (q^2)
$
\\
$\rho' \ra a_1$&$ \displaystyle
\braa{m_{\rho'}^2-m_{a_1}^2}\braa{\frac{r_{a_1}}{\sqrt{2}f_\pi}\sin \theta_\rho}\frac{ f_\pi}{q^2}
-\frac{g}{\sqrt{2}}\sin\theta_\rho \frac{g_{a_1}}{m_{a_1}^2}
\frac{m_{\rho'}^2-m_{a_1}^2}{m_{a_1}^2}
D^{a_1} (q^2)
$&$ \displaystyle
\sqrt{2} g\sin\theta_\rho \frac{g_{a_1}}{m_{a_1}^2}
D^{a_1} (q^2)
$&$ \displaystyle
0
$
\\
$h_1 \ra \rho$
&$ \displaystyle
-
\braa{m_{h_1}^2-m_{\rho}^2}\braa{-\frac{r_{a_1}}{\sqrt{2}f_\pi}\sin \theta_\rho}\frac{ f_\pi}{q^2}
-\frac{g}{\sqrt{2}}\sin\theta_\rho \frac{g_{a_1}}{m_{a_1}^2}
\frac{m_{h_1}^2-m_{\rho}^2}{m_{a_1}^2}
D^{a_1} (q^2)
$&$ \displaystyle
\sqrt{2} g\sin\theta_\rho \frac{g_{a_1}}{m_{a_1}^2}
D^{a_1} (q^2)
$&$ \displaystyle
-\frac{ g}{\sqrt{2}}\sin\theta_\rho \frac{g_{a_1}}{m_{a_1}^2}
D^{a_1} (q^2)
$
\\
$\rho' \ra h_1$&$ \displaystyle
\braa{m_{\rho'}^2-m_{h_1}^2}\braa{\frac{r_{a_1}}{\sqrt{2}f_\pi}\cos \theta_\rho}\frac{ f_\pi}{q^2}
-\frac{g}{\sqrt{2}}\cos\theta_\rho \frac{g_{a_1}}{m_{a_1}^2}
\frac{m_{\rho'}^2-m_{h_1}^2}{m_{a_1}^2}
D^{a_1} (q^2)
$&$ \displaystyle
\sqrt{2} g\cos\theta_\rho \frac{g_{a_1}}{m_{a_1}^2}
D^{a_1} (q^2)
$&$ \displaystyle
-\frac{ g}{\sqrt{2}}\cos\theta_\rho \frac{g_{a_1}}{m_{a_1}^2}
D^{a_1} (q^2)
$
\\
$b_1 \ra \omega$&$ \displaystyle
-\braa{m_{b_1}^2-m_{\omega}^2}\braa{-\frac{r_{a_1}}{\sqrt{2}f_\pi}\sin \theta_\omega}\frac{ f_\pi}{q^2}
-\frac{g}{\sqrt{2}}\sin\theta_\omega \frac{g_{a_1}}{m_{a_1}^2}
\frac{m_{b_1}^2-m_{\omega}^2}{m_{a_1}^2}
D^{a_1} (q^2)
$&$ \displaystyle
\sqrt{2} g\sin\theta_\omega \frac{g_{a_1}}{m_{a_1}^2}
D^{a_1} (q^2)
$&$ \displaystyle
-\frac{g}{\sqrt{2}}\sin\theta_\omega \frac{g_{a_1}}{m_{a_1}^2}
D^{a_1} (q^2)
$
\\
$\omega' \ra b_1$&$ \displaystyle
\braa{m_{\omega'}^2-m_{b_1}^2}\braa{\frac{r_{a_1}}{\sqrt{2}f_\pi}\cos \theta_\omega}\frac{ f_\pi}{q^2}
-\frac{g}{\sqrt{2}}\cos\theta_\omega \frac{g_{a_1}}{m_{a_1}^2}
\frac{m_{\omega'}^2-m_{b_1}^2}{m_{a_1}^2}
D^{a_1} (q^2)
$&$ \displaystyle
\sqrt{2} g\cos\theta_\omega \frac{g_{a_1}}{m_{a_1}^2}
D^{a_1} (q^2)
$&$ \displaystyle
-\frac{g}{\sqrt{2}}\cos\theta_\omega \frac{g_{a_1}}{m_{a_1}^2}
D^{a_1} (q^2)
$
\\\hline\hline
\end{tabular}
\end{center}
\label{Table:axial form factor}
\end{table*}

\begin{table}
\caption[]{ One pion and axial couplings  in the SU(4) HLS model. 
}
\begin{center}
\begin{tabular}{cc|cccccc}\hline\hline
$V_1$&$V_2$&$G_{V_1 V_2 \pi}^{(1)}$&$g_1(0)$
\\\hline
$a_1$&$ \rho$&
$ \displaystyle
\braa{m_{a_1}^2-m_{\rho}^2}\braa{\frac{r_{a_1}}{\sqrt{2}f_\pi}\cos \theta_\rho}
$&
$ \displaystyle
\sqrt{2} g\cos\theta_\rho \frac{g_{a_1}}{m_{a_1}^2}
$
\\
$\rho' $&$ a_1$&$ \displaystyle
\braa{m_{\rho'}^2-m_{a_1}^2}\braa{\frac{r_{a_1}}{\sqrt{2}f_\pi}\sin \theta_\rho}
$&$ \displaystyle
\sqrt{2} g\sin\theta_\rho \frac{g_{a_1}}{m_{a_1}^2}
$
\\
$h_1 $&$ \rho$
&$ \displaystyle
\braa{m_{h_1}^2-m_{\rho}^2}\braa{\frac{r_{a_1}}{\sqrt{2}f_\pi}\sin \theta_\rho}
$&$ \displaystyle
\sqrt{2} g\sin\theta_\rho \frac{g_{a_1}}{m_{a_1}^2}
$
\\
$\rho' $&$ h_1$&$ \displaystyle
\braa{m_{\rho'}^2-m_{h_1}^2}\braa{\frac{r_{a_1}}{\sqrt{2}f_\pi}\cos \theta_\rho}
$&$ \displaystyle
\sqrt{2} g\cos\theta_\rho \frac{g_{a_1}}{m_{a_1}^2}
$
\\
$b_1 $&$ \omega$&$ \displaystyle
\braa{m_{b_1}^2-m_{\omega}^2}\braa{\frac{r_{a_1}}{\sqrt{2}f_\pi}\sin \theta_\omega}
$&$ \displaystyle
\sqrt{2} g\sin\theta_\omega \frac{g_{a_1}}{m_{a_1}^2}
$
\\
$\omega' $&$ b_1$&$ \displaystyle
\braa{m_{\omega'}^2-m_{b_1}^2}\braa{\frac{r_{a_1}}{\sqrt{2}f_\pi}\cos \theta_\omega}
$&$ \displaystyle
\sqrt{2}g\cos\theta_\omega \frac{g_{a_1}}{m_{a_1}^2}
$
\\\hline\hline
\end{tabular}
\end{center}
\label{Table:coupling VVpi}
\end{table}

\section{Relations among one-pion decays of spin-1 mesons}
\label{sec:one pion decays}
In this section we give several relations among 
one-pion interactions of two spin-1 
mesons.

We would like to stress that all the one-pion decays of spin-1 mesons are expressed by one parameter $r_{a_1}/f_\pi$ reflecting the existence of the 
$\mbox{SU(4)}$ symmetry as shown in Eq.\,\eqref{GT relation}.
By using Eq\,\eqref{GT relation}, 
the one-pion decay widths of spin-1 mesons are easily calculated:
\bee{
\Gamma (V_i \ra V_f \pi)
=&
\frac{1}{8\kappa_{V_i \ra V_f }}\frac{\brad{\vec{p}_{V_i\ra V_f}}}{\pi m_{V_i}^2}
\braa{
g_{V_i V_f \pi}
}^2
\nn\\
&
\times\braa{
m_{V_i}^2
-
m_{V_f}^2
}^2
\braa{3+
\frac{\brad{\vec{p}_{V_i\ra V_f}}^2}{m_{V_f}^2}
}
\ ,
}
where the momentum is given as
\bee{
\brad{\vec{p}_{V_i \ra V_f}}
\equiv
\frac{1}{2}
\sqrt{m_{V_i}^2 - 2\braa{m_{V_f}^2 + m_{\pi}^2}+\frac{\braa{m_{V_f}^2- m_\pi^2}^2}{m_{V_i}^2}}
\ ,
}
and the factor $\kappa$ depends on the isospin of the initial and final states:
\bee{
\kappa_{h_1 \ra \rho} =& \kappa_{\omega' \ra b_1} = 1 
\ ,\nn\\
\kappa_{a_1 \ra \rho} =& \kappa_{\rho' \ra a_1} = \frac{3}{2}
\ ,\nn\\
\kappa_{b_1 \ra \omega} =& \kappa_{\rho' \ra h_1} = 3 
\ .
}
The unknown parameter $r_{a_1}/f_\pi$ in the coupling $g_{VV\pi}$ is canceled by taking ratios of these decay widths:
\bee{
&
\frac{\Gamma (\rho' \ra h_1 \pi)}{\Gamma (a_1 \ra \rho \pi)}
=
0.16 \pm 0.07
\ ,
\nn\\
&
\frac{\Gamma (h_1 \ra \rho \pi)}{\Gamma (\rho' \ra a_1 \pi)}
=
6.4\pm4.3
\ ,
\nn\\
&
\frac{\Gamma (\rho' \ra a_1 \pi)}{\Gamma (a_1 \ra \rho \pi)}
=
\braa{
0.15\pm0.11
} \tan^2\theta_\rho
\ ,
\nn\\
&
\frac{\Gamma (h_1 \ra \rho \pi)}{\Gamma (a_1 \ra \rho \pi)}
=
\braa{
1.0\pm0.3
} \tan^2\theta_\rho
\ ,
\nn\\
&
\frac{\Gamma (b_1 \ra \omega \pi)}{\Gamma (\omega' \ra b_1 \pi)}
=
\braa{
3.9\pm1.9
} \tan^2\theta_\omega
\ ,
\label{ratio between the widths}
}
where the numerical factors in the RHS are simply evaluated from the corresponding kinematical factors calculated by using the masses listed in Table\,\ref{Table:masses}. 
Errors in the RHS are estimated  from the errors listed in the table.
Since the first two relations are independent of the parameters, 
experimental measurements of these ratios will check  the existence of the $\mbox{SU(4)}$ symmetry. 
Then, we can determine the mixing angles from the latter three relations.

\section{Extended KSRF relations}
\label{sec:KSRF}

In this section, we  
derive the Kawarabayashi-Suzuki-Riazuddin-Fayyazuddin (KSRF) relations
among the $\rho$ meson mass, the $\rho\pi\pi$ coupling and the $\rho$-photon mixing strength, as well as their extension to the $\rho'$ meson.

The interactions among one gauge field and two pion fields are included in 
the $\mc{L}_V$ of the Lagrangian\,\eqref{Lagrangian full2}.
Similarly to the one-pion interactions studied in the previous section, due to the existence of the $a_1$-$\pi$ mixing, the three point interaction ${\mathcal L}_{\rm int}^{(3)}$ generates the interactions among a spin-1 meson and two pions.
The resultant 
 effective vertices  among two pions and one vector meson are given by
\bee{
&\Gamma^\mu \brac{\rho_\mu^a(p),  \pi^b (p_1) , \pi^c (p_2)} 
\nn\\
=&
 -i\epsilon^{abc}\brac{g^{(T)}_{\rho \pi\pi}(p^2)P^{\mu\nu} (p) \brac{p_1-p_2}_\nu 
+ g^{(L)}_{\rho \pi\pi}(p_1 ,p_2){p}^\mu}
\ ,
\nn\\
&\Gamma^\mu \brac{\braa{\rho'}_\mu^a(p),  \pi^b (p_1) , \pi^c (p_2)} 
\nn\\
=&
 -i\epsilon^{abc}\brac{g^{(T)}_{\rho' \pi\pi}(p^2)P^{\mu\nu} (p) \brac{p_1-p_2}_\nu
+ g^{(L)}_{\rho' \pi\pi}(p_1 ,p_2){p}^\mu
}
}
where $p=p_1+p_2$,
and
\bee{
g^{(T)}_{\rho \pi\pi}(p^2)=&\frac{m_\rho^2 -r_{a_1}^2 p^2}{\sqrt{2}gf_\pi^2}\cos \theta_\rho
\ ,
\nn\\
g^{(T)}_{\rho' \pi\pi}(p^2)=&\frac{m_{\rho'}^2 -r_{a_1}^2 p^2}{\sqrt{2}gf_\pi^2}\sin \theta_\rho
\ ,
\nn\\
g^{(L)}_{\rho \pi\pi}(p_1 ,p_2)=&\frac{m_\rho^2\cos \theta_\rho}{\sqrt{2}gf_\pi^2}
\frac{\braa{p_2}^2-\braa{p_1}^2}{\braa{p_1 + p_2}^2}
\ ,
\nn\\
g^{(L)}_{\rho' \pi\pi}(p_1 ,p_2)=&\frac{m_{\rho'}^2\sin \theta_\rho}{\sqrt{2}gf_\pi^2}
\frac{\braa{p_2}^2-\braa{p_1}^2}{\braa{p_1 + p_2}^2}
\label{rho pi pi coupling}
\ .
}
Since $g_{V\pi\pi}^{(L)}(p_1,p_2)$ ($V=\rho\,,\,\rho'$) vanishes for on-shell pion, only $g_{V\pi\pi}^{(T)}(p^2)$ is relevant for $V\to \pi\pi$ decay and the electromagnetic form factor of pion.
Furthermore, when the vector mesons are on their mass shell, the ratio of two $V\pi\pi$ couplings is related to the mixing angle as
\begin{equation}
\frac{g_{\rho'\pi\pi}^{(T)}(m_{\rho'}^2) }{ g_{\rho\pi\pi}^{(T)}(m_{\rho}^2) } = \frac{ m_{\rho'}^2 }{m_\rho^2} \, \tan\theta_\rho \ .
\label{ratio Vpp}
\end{equation}

We introduce the photon field $A_\mu$ by replacing the external gauge field as
\begin{equation}
{\mathcal V}_\mu = e A_\mu Q \ ,
\end{equation}
where $e$ is the electromagnetic coupling constant, and
\begin{equation}
Q = \sqrt{2}\braa{S^3+\frac{1}{3}S^0}
=\braa{\mt{t^3 + \frac{1}{6}1_2 &0\\0&t^3 + \frac{1}{6}1_2 }}
\ .
\end{equation}
The $\mc{L}_V$ part of the Lagrangian generates the mixing between a vector meson and the photon.
The mixing strengths for $\rho$, $\rho'$, $\omega$ and $\omega'$ mesons are expressed as 
\begin{align}
g_{\rho}\equiv&
\frac{\sqrt{2}m_{\rho}^2\cos \theta_\rho}{g} 
\ ,~~~~
g_{\rho'}\equiv
\frac{\sqrt{2}m_{\rho'}^2\sin \theta_\rho}{g} 
\ ,
\label{coupling grho} \\
g_{\omega}\equiv&
\frac{\sqrt{2}m_{\omega}^2\cos \theta_\omega}{3g_B} 
\ ,~~~~
g_{\omega'}\equiv
\frac{\sqrt{2}m_{\omega'}^2\sin \theta_\omega}{3g_B} 
\ .
\label{coupling gomega}
\end{align}
Similarly to Eq.\,(\ref{ratio Vpp}), several ratios of two of above quantities are expressed as
\begin{align}
\frac{g_{\rho'} }{ g_\rho } = & \frac{ m_{\rho'}^2 }{m_\rho^2} \, \tan\theta_\rho \ , \notag\\
\frac{g_{\omega} }{ g_\rho } = & \frac{1}{3}\, \frac{ m_{\omega}^2 }{m_\rho^2} \frac{g}{g_B}\frac{\cos\theta_\omega}{\cos\theta_\rho}\ , \notag\\
\frac{g_{\omega'} }{ g_\omega } = & \frac{ m_{\omega'}^2 }{ m_\omega^2} \, \tan\theta_\omega \ , 
\label{ratio Vg}
\end{align}

Now, comparing the expressions in Eq.\,(\ref{coupling grho}) with the two-pion vertices in Eq.\,(\ref{rho pi pi coupling}),
one
can find the KSRF I relation and the extended one for $\rho'$ in the soft momentum limit, $p =0$:
\bee{
g_\rho=2 g_{\rho\pi\pi}^{(T)}(p^2=0) f_\pi^2
\ ,~~~
g_{\rho'}=2 g_{\rho' \pi\pi}^{(T)}(p^2=0) f_\pi^2
\ .
}
On the other hand, for the on-shell vector mesons, they become
\bee{
\frac{2 g_{\rho\pi\pi}^{(T)}(m_\rho^2) f_\pi^2}{g_\rho}
=\frac{2 g_{\rho'\pi\pi}^{(T)}(m_{\rho'}^2) f_\pi^2}{g_{\rho'}}=1-r_{a_1}^2
\label{KSRF I on-shell}
\ .
}
This implies that the deviation for the on-shell $\rho$ from the KSRF I relation is caused by the term including $r_{a_1}$, which is generated from $\mc{L}_{\rm int}^{(3)}$, in $g_{\rho\pi\pi}^{(T)}(p^2)$.

Let us consider the relations among several relevant decay widths. 
The decay widths for the $\rho \to \pi\pi$ and $\rho^0 \to e^+e^-$ 
 are calculated as
\bee{
\Gamma \braa{\rho \ra \pi\pi}
=&
\frac{1}{6\pi m_\rho^2}\brac{\frac{m_\rho^2 -4m_\pi^2}{4}}^{\frac{3}{2}}
\brad{g^{(T)}_{\rho \pi\pi}(m_\rho^2)}^2
\ ,
\label{Decay width rho to pipi}
\\
\Gamma \braa{\rho^0 \ra e^+e^-}
=&
\frac{4\pi \alpha^2}{3}
\brad{\frac{g_{\rho}}{m_\rho^2}}^2
\frac{m_\rho^2 + 2m_e^2}{m_\rho^2}\sqrt{m_\rho^2-4m_e^2}
\label{Decay width rho to ee}
\ ,
}
and similarly for $\rho' \to \pi\pi$ and ${\rho '}, \omega, \omega' \to e^+ e^-$. 
Combining the relations in Eq.\,(\ref{ratio Vpp}) and (\ref{ratio Vg}), we obtain
\bee{
\frac{\Gamma \braa{\rho' \ra \pi\pi}}
{\Gamma \braa{\rho \ra \pi\pi}}
=&
\braa{
28 \pm 2
}
\tan^2 \theta_\rho
\ ,
\label{rho pipi decay}\\
\frac{\Gamma \braa{{\rho'}^0 \ra e^+e^-}}
{\Gamma \braa{\rho^0 \ra e^+e^-}}
=&
\braa{
1.89\pm 0.03
}
\tan^2 \theta_\rho
\ ,
\label{rhoprime decays} \\
\frac{ \Gamma \left( \omega \to e^+ e^- \right) }{ \Gamma \left( \rho^0 \to e^+ e^- \right)  } = & \left( 
0.112 \pm 0.000
\right) \, \frac{g^2}{g_B^2}\frac{ \cos^2\theta_\omega }{ \cos^2 \theta_\rho } \ ,
\label{omega ee decay} \\
\frac{ \Gamma \left( \omega' \to e^+ e^- \right) }{ \Gamma \left( \omega \to e^+ e^- \right)  } = & \left( 
1.01\pm 0.03
\right) \,  \tan^2\theta_\omega  \ ,
\label{omegap ee decay}
}
where $0.000$ in the third equations implies that the error is smaller than $0.0005$. 
Taking the ratio of Eq.~(\ref{rho pipi decay}) and Eq.~(\ref{rhoprime decays}), we obtain the following parameter free relation:
\begin{equation}
\frac{ \Gamma \left( \rho' \to \pi\pi \right) }{ \Gamma \left( \rho \to \pi\pi \right) }
\frac{ \Gamma \left( \rho^0 \to e^+e^- \right) }{ \Gamma \left( \rho^{\prime0} \to e^+e^- \right) } = \left(
15 \pm 1
\right) \ ,
\end{equation}
which is regarded as an experimental check of the existence of $\mbox{SU(4)}$ symmetry for spin-1 mesons.

The relations in Eqs.\,(\ref{rho pipi decay})-(\ref{omegap ee decay}) can be  
used to determine the relevant model parameters.
At this moment, we can set up an upper limit for the mixing angle $\tan^2\theta_\rho$ in the following way: The 
total decay width of $\rho'$ and the partial decay width of the $\rho \ra \pi\pi$ channel are known as $\Gamma^{\rm total} (\rho') = 400\pm 60\,\mbox{MeV}$ and $\Gamma(\rho \ra \pi\pi) = 147.8\,\mbox{MeV}$, respectively.
Then the upper limit of the LHS of Eq.~(\ref{rho pipi decay}) is estimated as
\begin{equation}
\frac{ \Gamma \left( \rho' \to \pi\pi \right) }{ \Gamma \left( \rho \to \pi\pi \right) } \le
\frac{ \Gamma^{\rm total} \left( \rho' \right) }{ \Gamma \left( \rho \to \pi\pi \right) }  \sim  3 \ .
\end{equation}
From this together with Eq.~(\ref{rho pipi decay}), 
the limit is obtained as
\bee{
\tan^2\theta_\rho \lesssim 0.1
\label{upper limit of tan2}
\ ,
}
which implies that the mixing between 
$V_{\parallel}$ and $V_{\perp(1)}$
 is not large.
From this upper limit for the mixing angle, the ratio of $e^+e^-$ decays of $\rho'$ and $\rho$ mesons has an upper limit as
\begin{equation}
\frac{\Gamma \braa{{\rho'}^0 \ra e^+e^-}}
{\Gamma \braa{\rho^0 \ra e^+e^-}} \lesssim 0.2 \ .
\end{equation}
Using the upper limit for $\tan^2\theta_\rho$ in Eqs.\,\eqref{upper limit of tan2}, we obtain the upper limits for the ratios of one-pion decays of spin-1 mesons:
\begin{align}
& \frac{\Gamma (\rho' \ra a_1 \pi)}{\Gamma (a_1 \ra \rho \pi)} \lesssim 0.02 \ ,
\notag\\
& \frac{\Gamma (h_1 \to \rho \pi)}{\Gamma (a_1 \to \rho \pi)} \lesssim 0.1 \ ,
\notag\\
\end{align}
which can be tested in future experiments.

Next, we consider a constraint from Eq.~(\ref{omega ee decay}). 
Using the experimental value $\Gamma(\omega \to e^+ e^-) = 0.60 \pm 0.02\,\mbox{keV}$, we obtain 
\begin{equation}
\frac{g^2}{g_B^2}\frac{\cos^2\theta_\omega}{\cos^2\theta_\rho} =
0.76 \pm 0.03
\label{ratio cos omega}
\ .
\end{equation}
When the gauge couplings for $\mbox{SU(4)}$ HLS and $\mbox{U(1)}$ HLS are equal to each other, this together with Eq.~(\ref{upper limit of tan2}) gives a constraint as 
\begin{equation}
\tan^2 \theta_\omega \lesssim 0.5
\ ,
\label{upper limit theta omega}
\end{equation}
and 
\bee{
& \frac{\Gamma (b_1 \ra \omega \pi)}{\Gamma (\omega' \ra b_1 \pi)} \lesssim 2 \ ,
\nn\\
& 
\frac{ \Gamma \left( \omega' \to e^+ e^- \right) }{ \Gamma \left( \omega \to e^+ e^- \right)  } \lesssim 0.5 \ .
}

\section{Numerical analysis}
\label{sec: numerical analysis}

In this section, we determine the model parameters from the relevant experimental data, and make several phenomenological predictions.

We first construct the electromagnetic form factor of pion.
From the second term in the last line of Eq.\,\eqref{Lagrangian including alpha}, we can read the
direct $\gamma \pi \pi $ coupling as
\bee{
g_{\gamma \pi \pi}\equiv&
1-\frac{m_{\rho}^2\cos^2 \theta_\rho+m_{\rho'}^2\sin^2 \theta_\rho}{g^2f_\pi^2} \ .
\label{direct gamma pi pi}
}
From this and Eqs.\,(\ref{rho pi pi coupling}) and (\ref{coupling grho}), the 
pion space-like form factor is given by
\bee{
F^{\pi^\pm}_V (Q^2)
=&
g_{\gamma\pi\pi}
+ 
\frac{g_\rho g_{\rho \pi\pi}^{(T)} (-Q^2)}{m_\rho^2 +Q^2}
+ 
\frac{g_{\rho'} g_{\rho' \pi\pi}^{(T)} (-Q^2)}{m_{\rho'}^2 +Q^2}
\label{pion form factor}
}
with $Q^2 = -q^2$, where $q$ is the photon momentum.
This form factor is normalized as $F^{\pi^\pm}_V (Q^2=0) =1$
reflecting the existence of  the electromagnetic U(1) symmetry.  
From this, the  pion charge radius is calculated as 
\bee{
\brae{r^2}^{\pi^\pm}_V\equiv& 
-6\left.
\frac{\pd F^{\pi^\pm}_V (Q^2)
}{\pd Q^2}
\right|_{Q^2=0}
=
6
\frac{1-r_{a_1}^2}{g^2 f_\pi^2} 
\label{pion charge radius}
\ .
}

By using Eqs.\,\eqref{coupling grho}, 
\eqref{KSRF I on-shell}, and
\eqref{pion charge radius},
the parameters are expressed as 
\bee{
\brad{r_{a_1} }=& \sqrt{1- \braa{\frac{2 g_{\rho\pi\pi}^{(T)}(m_\rho^2) f_\pi^2}{g_\rho}}}
\ ,\label{parameter ra1}\\
\brad{g}
=&
\frac{m_\rho}{f_\pi}\sqrt{
\braa{\frac{2g_{\rho\pi\pi}^{(T)}(m_\rho^2)f_\pi^2}{g_\rho}}
/
\braa{\frac{1}{6}\brae{r^2}^{\pi^\pm}_V m_\rho^2}
}
\ ,\label{parameter g}\\
\tan^2 \theta_\rho
=&
\frac{\braa{\frac{1}{6}\brae{r^2}^{\pi^\pm}_V m_\rho^2}}{\braa{\frac{2g_{\rho\pi\pi}^{(T)}(m_\rho^2) f_\pi^2}{g_\rho}}
\braa{\frac{g_\rho}{\sqrt{2}m_\rho f_\pi}}^2
}
-1
\ .
\label{parameters expressed by grhopipi grho r2}
}
We should note that, since the pion charge radius is positive and
\begin{equation}
\frac{2 g_{\rho\pi\pi}^{(T)}(m_\rho^2) f_\pi^2}{g_\rho}=\frac{g^2 f_\pi^2}{6 \brae{r^2}^{\pi^\pm}_V}
\end{equation}
is satisfied, one can find that the couplings $g_{\rho\pi\pi}^{(T)}(m_\rho^2)$ and $g_{\rho}$ have the same sign.  
Then, the inside of the square root in the RHS of  Eq.~(\ref{parameter g})  is always positive.

\begin{figure}[h]
\centering	\includegraphics[width=90mm]{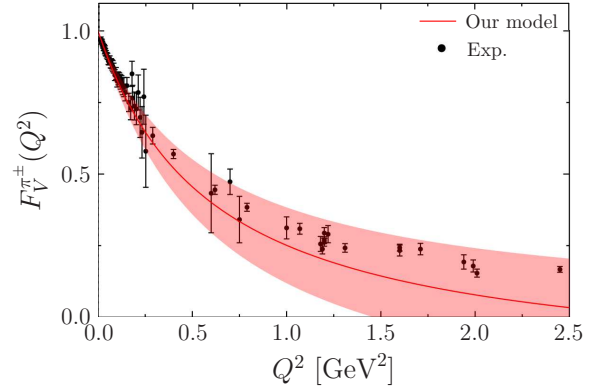}
\caption{Electromagnetic form factor of pion.  Predicted form factor from the central values of the parameters in Eq.~(\ref{values of parameters}) is shown by the red curve.  Shaded area shows the errors of the parameters.  Black 
 dots express the experimental data given in Refs.\,\cite{Amendolia:1986wj,Bebek:1977pe,
Volmer:2000ek,Horn:2006tm,Tadevosyan:2007yd}.}
\label{vector pion form factor}
\end{figure}

Substituting the experimental values  listed in Table\,\ref{Table:Experimental values}
into Eqs.~(\ref{parameter ra1})-(\ref{parameters expressed by grhopipi grho r2}), we have
\bee{
\brad{r_{a_1} }=& 
0.41 \pm 0.10
\ ,\nn\\
\brad{g}
=&
7.1 \pm 0.5
\ ,\nn\\
\tan^2 \theta_\rho
=&
-0.03 \pm 0.25
\label{values of parameters}
}
where we added
10\% errors  expected from  higher order corrections~\cite{Harada:2003jx}. 
Then the upper limit of $\tan^2\theta_\rho$, 
$\tan^2\theta_\rho \lesssim 0.1$ given in Eq.\,\eqref{upper limit of tan2},
 is within the errors of above determination. 
The electromagnetic form factor obtained from these values is 
shown in Fig.\,\ref{vector pion form factor} together with the experimental data. 
This shows that the predicted form factor reasonably reproduce
 the experimental data, taking account of their errors.

We would like to note that the both numerators of the $\rho$ and $\rho'$ contributions in Eq.\,\eqref{pion form factor} have the same sign, which is contrasted to the result by a holographic QCD model~\cite{Harada:2010cn}.
Furthermore, the direct $\gamma\pi\pi$ coupling is evaluated through
\bee{
g_{\gamma\pi\pi}
=&
1-
\frac{m_{\rho'}^2}{m_\rho^2}
\frac{\braa{\frac{1}{6}\brae{r^2}^{\pi^\pm}_V m_\rho^2}}{\braa{\frac{2g_{\rho\pi\pi}^{(T)}(m_\rho^2) f_\pi^2}{g_\rho}}
}
+
\braa{\frac{g_\rho}{\sqrt{2}m_\rho f_\pi}}^2
\braa{\frac{m_{\rho'}^2}{m_\rho^2}-1}
\ ,
}
as $g_{\gamma\pi\pi}=
-0.28 \pm 0.13$ by using
the values in Table\,\ref{Table:Experimental values}.
This result means that there is a slight  deviation from the vector meson dominance.

At the end of this section, we estimate several decay widths of spin-1 mesons by using the parameter set given in Eqs.\,\eqref{values of parameters} and show them in Table\,\ref{Table:decay widths from numerical analysis}. 
\begin{table}
\caption[]{ Predicted values of one-pion decay widths of spin-1 mesons estimated  from the parameter set in Eq.\,\eqref{values of parameters}.}
\begin{center}
\begin{tabular}{c|c}\hline\hline
Decay mode & Partial width (MeV)
\\\hline
$\Gamma \braa{a_1 \ra \rho\pi}$ &
$470 \pm 400$
\\
$\Gamma \braa{\rho' \ra a_1\pi}$ & $< 50$
\\
$\Gamma \braa{h_1 \ra \rho\pi}$ & $<60$ 
\\
$\Gamma \braa{\rho' \ra h_1\pi}$ & $80 \pm 300$
\\\hline\hline
\end{tabular}
\end{center}
\label{Table:decay widths from numerical analysis}
\end{table}

\section{Summary and Discussions}
\label{sec:summary}

We constructed a chiral Lagrangian with an $\mbox{SU(4)}\times\mbox{U(1)}$ hidden local symmetry which includes  the  spin-1 mesons,
$\braa{\rho, a_1, \rho', \omega', b_1, f_1, h_1}$, together with pion.
 We  found  
that each coupling of the interaction among one pion and two spin-1 mesons is proportional to the mass difference of the relevant spin-1 mesons similarly to the Goldberger-Treiman relation. 
In addition, there were
the relations among  one-pion  decays of spin-1 mesons  thanks to the existence of the SU(4) emergent symmetry. 
Furthermore, we found a relation among the mass of $\rho'$ meson, the $\rho'\pi\pi$ coupling and the $\rho'$-photon mixing strength as well as the Kawarabayashi-Suzuki-Riazuddin-Fayyazuddin relation for the $\rho$ meson. 
We summarize these predictions in Table\,\ref{Table:Predictions}.
\begin{table}
\begin{center}
\caption[]{
Predictions obtained from the $\mbox{SU(4)}\times\mbox{U(1)}$ HLS model.
}
\label{Table:Predictions}
\begin{tabular}{cccccccc}\hline\hline
\multicolumn{3}{c}{Independent of the parameters}
&
\\\hline
$\displaystyle \frac{\Gamma (\rho' \ra h_1 \pi)}{\Gamma (a_1 \ra \rho \pi)}$
&$=$
&$0.16 \pm 0.07$
&
\\
$\displaystyle \frac{\Gamma (h_1 \ra \rho \pi)}{\Gamma (\rho' \ra a_1 \pi)}$
&$=$
&$6.4\pm4.3$
&
\\\hline\hline
\multicolumn{3}{c}{Dependent on $\theta_\rho$ or $\theta_\omega$}
\\\hline
$\displaystyle \frac{\Gamma (\rho' \ra a_1 \pi)}{\Gamma (a_1 \ra \rho \pi)}$
&$=$
&$\braa{0.15\pm0.11}\, \tan^2\theta_\rho$
&$\lesssim$&
$0.015$
\\
$\displaystyle \frac{\Gamma (h_1 \ra \rho \pi)}{\Gamma (a_1 \ra \rho \pi)}$
&$=$
&$\braa{1.0\pm0.3}\, \tan^2\theta_\rho$
&$\lesssim$&
$0.10$
\\
$\displaystyle \frac{\Gamma \braa{\rho' \ra \pi\pi}}{ {\Gamma \braa{\rho \ra \pi\pi}}}$
&$=$&
$\braa{28 \pm 2}\tan^2 \theta_\rho$
&&(Input)
\\
$\displaystyle \frac{\Gamma \braa{{\rho'}^0 \ra e^+e^-}}{\Gamma \braa{\rho^0 \ra e^+e^-}}$
&$=$&
$\braa{1.89\pm 0.03}\tan^2 \theta_\rho$
&$\lesssim$&
$0.2$
\\
$\displaystyle \frac{\Gamma \left( \omega \to e^+ e^- \right) }{ \Gamma \left( \rho^0 \to e^+ e^- \right)} $  &$=$
&$\left(0.11219 \pm 0.00004\right) \, 
\displaystyle \frac{ g^2 }{ g_B^2} \frac{ \cos^2\theta_\omega }{ \cos^2 \theta_\rho }$
&&(Input)
\\
$\displaystyle \frac{\Gamma (b_1 \ra \omega \pi)}{\Gamma (\omega' \ra b_1 \pi)}$
&$=$
&$\braa{3.9\pm1.9}\, \tan^2\theta_\omega$
\\
$\displaystyle \frac{\Gamma \left( \omega' \to e^+ e^- \right) }{ \Gamma \left( \omega \to e^+ e^- \right)}$ &$=$&
$\left( 1.01\pm 0.03\right) \, \displaystyle \tan^2\theta_\omega $
\\\hline\hline
\end{tabular}
\end{center}
\end{table}

Using two ratios indicated in Table\,\ref{Table:Predictions} together with the total widths of $\rho'$ and $\omega'$ mesons, we obtained several upper limit for the ratios as shown in the last column of the table. 
By testing them in future experiments, we can verify the existence of the emergent symmetry.

There also exist hadronic decays which involve the intrinsic parity odd terms shown 
 in Appendix\,\ref{sec:Intrinsic parity odd terms}.
Possible decay 
 modes are expressed by ``$\times$'' in 
Table\,\ref{Table:Decay channel}, 
such as $\rho' \ra \omega \pi$ and $\omega \ra 3\pi$.
We listed the allowed operators in Appendix\,\ref{sec:Intrinsic parity odd terms}.

Similarly to the result obtained in the generalized HLS at ${\mathcal O}(p^2)$~\cite{Bando:1987ym,Bando:1987br}, 
we also found $\Gamma\braa{a_1 \ra \pi \gamma}=0$ together with $\Gamma\braa{b_1 \ra \pi \gamma}=0$ and $\Gamma\braa{h_1 \ra \pi \gamma}=0$  at the leading order, 
 as we showed some  detail calculations  in Appendix\,\ref{sec:to pi gamma}.
As in the case of the generalized HLS, we expect that non-vanishing contributions will be produced 
 by higher order correction of the derivative expansion\,\cite{Bando:1987ym,Bando:1987br,Nagahiro:2008cv}.

We have to remark that our analysis are done in the chiral broken phase since we used the nonlinear realization  of the chiral symmetry. 
On the other hand, the existence of the emergent symmetry is proposed 
by reducing the Dirac zero mode in the lattice QCD, which corresponds to
remove the dominant contribution of the chiral symmetry breaking
as shown by the Banks-Casher relation.
 It will be interesting to clarify  the correspondence between our model and the lattice QCD result,  which we leave for future works.

\begin{table}
\begin{center}
\caption[]{Two-body pionic decay channels. 
``$-$'' implies that the decay mode is prohibited kinematically or by the isospin, $\mc{P}$, and $\mc{C}$ while 
``$\times$'' and ``$\sqrt{}$'' mean that such decay channel is allowed through the operators with the intrinsic parity odd and even, respectively. 
}
\begin{tabular}{cc|ccccccccc|ccc}\hline\hline
Initial &mass& \multicolumn{10}{|c}{channel}
\\
&(MeV)&
$\pi \pi$&$\rho\pi$&$\omega\pi$&$\rho'\pi$&$\omega'\pi$&$a_1 \pi$&$f_1\pi$&$b_1\pi$&$h_1\pi$&$3\pi$&$\eta \pi\pi$&$4\pi$
\\\hline
$\rho$&$775.26$&$\sqrt{}$
&$-$&$-$&$-$&$-$&$-$&$-$&$-$&$-$
&$-$&$-$&$\sqrt{}$
\\
$\omega$&$782.65$&
$-$&$-$&$-$&$-$&$-$&$-$&$-$&$-$&$-$
&$\times$&$-$&$-$
\\
$\rho'$&$1465$&
$\sqrt{}$
&$-$
&$\times$&$-$&$-$&$\sqrt{}$&$-$&$-$&$\sqrt{}$
&$-$&$\times$&$\sqrt{}$
\\
$\omega'$&$\sim1425$&
$-$&$\times$
&$-$
&$-$&$-$&$-$&$-$&$\sqrt{}$
&$-$&$\times$&$-$
&$-$
\\
$a_1$&$1230$&
$-$&$\sqrt{}$
&$-$&$-$&$-$&$-$&$-$&$-$&$-$
&$\sqrt{}$&$-$
&$-$
\\
$f_1$&$1281.9$&
$-$&$-$
&$-$&$-$&$-$&$-$&$-$&$-$&$-$
&$-$&$\sqrt{}$&$\times$
\\
$b_1$&$1229.5$&
$-$&$-$
&$\sqrt{}$&$-$&$-$&$-$&$-$&$-$&$-$
&$-$&$\sqrt{}$&$\times$
\\
$h_1$&$1170$&
$-$&$\sqrt{}$
&$-$&$-$&$-$&$-$&$-$&$-$&$-$
&$\sqrt{}$&$-$&$-$
\\\hline\hline
\end{tabular}
\end{center}
\label{Table:Decay channel}
\end{table}

\section*{ACKNOWLEDGEMENTS}
We would like to thank Yuichi Motohiro for useful discussions.
The work of MH is supported in part by the JSPS Grant-in-Aid for Scientific Research (c) No.~16K05345.
The work of HN is supported in part by JSPS KAKENHI Grant Number~JP16J03578.

\appendix

\section{Generators of an $\mbox{SU(4)}$}
\label{sec:Generators}

The generators of an $\mbox{SU(4)}$ are defined as
 $T^A = \{ S^a, X^a_{(3)}, X^a_{(1)}, X^a_{(2)}, X^0_{(3)}, X^0_{(1)}, X^0_{(2)} \}$ with $A = 1,\ldots,15$ and $a=1,2,3$:  
\bee{
S^a \equiv &\frac{1}{\sqrt{2}}\braa{\mt{t^a&0\\0&t^a}}
\ ,
\nn\\
X^a_{(3)} \equiv& \frac{1}{\sqrt{2}}\braa{\mt{t^a&0\\0&-t^a}}
\ ,
\nn\\
X^a_{(1)} \equiv &\frac{1}{\sqrt{2}}\braa{\mt{0&t^a\\t^a&0}}
\ ,
\nn\\
X^a_{(2)} \equiv&\frac{1}{\sqrt{2}}\braa{\mt{0&-it^a\\it^a&0}}
\ ,
\nn\\
X^0_{(3)}\equiv&\frac{1}{2\sqrt{2}}\braa{\mt{1_2&0\\0&-1_2}}
\ ,
\nn\\
X^0_{(1)} \equiv &\frac{1}{2\sqrt{2}}\braa{\mt{0&1_2\\1_2&0}}
=
\frac{1}{2\sqrt{2}} \bar{\Sigma}
\ ,
\nn\\
X^0_{(2)} \equiv&\frac{1}{2\sqrt{2}}\braa{\mt{0&-i1_2\\i1_2&0}}
}
where $a =1, 2, 3$ and $t^a =\sigma^a /2$.
An $\mbox{SU(4)}$ group includes its subgroup $\mbox{SU(2)}$ whose generator is $S^a$.
The generator of a $\mbox{U(1)}$ is also defined as
\bee{
S^0 \equiv\frac{1}{2\sqrt{2}}\braa{\mt{1_2&0\\0&1_2}}
\ .
}

The commutation relation between the generators are obtained as
\bee{
\brac{
X^a_{(i)},
X^b_{(j)}
}
=&
\frac{1}{\sqrt{2}}
i\epsilon^{ijk}\delta^{ab} X_{(k)}^0
+
\frac{1}{\sqrt{2}}
i\epsilon^{abc}\delta_{ij}
S^c
\ ,
\nn\\
\brac{
X^a_{(i)},
X^0_{(j)}
}
=&
\frac{1}{\sqrt{2}}
i\epsilon^{ijk} X_{(k)}^a
\ ,
\nn\\
\brac{
X^0_{(i)},
X^0_{(j)}
}
=&
\frac{1}{\sqrt{2}}
i\epsilon^{ijk} X_{(k)}^0
\ ,
\nn\\
\brac{
S^a,
X^0_{(i)}
}
=&
0
\ ,
\nn\\
\brac{
S^a,
X^b_{(i)}
}
=&
\frac{1}{\sqrt{2}}
i\epsilon^{abc}
X_{(i)}^c
\ ,
\nn\\
\brac{
S^a,
S^b
}
=&
\frac{1}{\sqrt{2}}
i\epsilon^{abc}
S^c
\ .
}

\begin{widetext}

\section{Detailed calculations  of Lagrangian terms  }
\label{sec:expansion}

In this  appendix 
we show detailed calculations for obtaining the interaction terms among the spin-1 mesons and pion.

We expand the Maurer-Cartan 1-forms given in Eq.\,\eqref{definition of alpha} in the unitary gauge $p=s=\tilde{s}=0$: 
\bee{
\hat{\alpha}_{\mu \parallel} (x) =&
-\frac{1}{2iF_\pi}
\brac{\pi , \pd_\mu \pi}
-
V_{\mu \parallel}
+
2\tr\brac{\Xi (\pi) \cdot \mc{V}_\mu\cdot \Xi^\dagger (\pi) \cdot S^a}S^a
+
\cdots
\ ,
\nn\\
\hat{\alpha}_{\mu \perp (3)} (x) =&
\frac{1}{F_\pi}\pd_\mu \pi
-
\frac{1}{6F_\pi^3}
\brac{\pi , \brac{\pi , \pd_\mu \pi}}
-
V_{\mu \perp (3)}
+
2\tr\brac{\Xi (\pi) \cdot \mc{V}_\mu\cdot \Xi^\dagger (\pi) \cdot X_{\perp(3)}^a}X_{\perp(3)}^a
+
\cdots
\ ,
\nn\\
\hat{\alpha}_{\mu\perp (1)}^{(m)} (x) =&V_{\mu \perp(1)}
\ ,~~
\hat{\alpha}_{\mu\perp (2)}^{(m)} (x) =V_{\mu \perp(2)}
\ ,~~
\hat{\alpha}_{\mu\perp (3)}^{(m)} (x) =V_{\mu \perp(3)}
\ .
\label{expansion of alpha}
}
with 
$V_{\mu \parallel} = 2\tr\brac{V_\mu \cdot S^a}S^a$ and
$V_{\mu \perp{(i)}} = 2\tr\brac{V_\mu \cdot X^a_{(i)}}X^a_{(i)}$.
By using Eq.\,\eqref{expansion of alpha}, ${\mc{L}}_{V}$ is written as
\bee{
 {\mc{L}}_{V}
=&
\bar{a}_{(1)}
F^2\tr\brac{
\braa{\frac{1}{F_\pi}\pd_\mu \pi
-\frac{1}{6F_\pi^3}
\brac{\pi , \brac{\pi , \pd_\mu \pi}}
+2\tr\brac{\Xi (\pi) \cdot \mc{V}_\mu\cdot \Xi^\dagger (\pi) \cdot X_{\perp(3)}^a}X_{\perp(3)}^a
}^2}
\nn\\
&
+
\bar{a}_{(2)}
F^2\tr\brac{
\braa{\frac{1}{2iF_\pi}\brac{\pi , \pd_\mu \pi}
-V_{\mu \parallel}
+2\tr\brac{\Xi (\pi) \cdot \mc{V}_\mu\cdot \Xi^\dagger (\pi) \cdot S^a}S^a}^2}
\nn\\
&
+
\bar{a}_{(3)}
F^2\tr\brac{\braa{V_{\mu \perp (1)} }^2}
+
\bar{a}_{(4)}
F^2\tr\brac{\braa{V_{\mu \perp (2)}}^2}
+
\bar{a}_{(5)}
F^2\tr\brac{\braa{V_{\mu \perp (3)}}^2}
\nn\\
&
+\bar{a}_{(6)}
2F^2\tr\brac{\braa{\frac{1}{2iF_\pi}\brac{\pi , \pd_\mu \pi}
-V_{\mu \parallel}
+2\tr\brac{\Xi (\pi) \cdot \mc{V}_\mu\cdot \Xi^\dagger (\pi) \cdot S^a}S^a}
\cdot  V^{\mu}_{\perp (1)} \cdot \bar{\Sigma}}
\nn\\
&
-
\bar{a}_{(7)}
2F^2\tr\brac{
\braa{\frac{1}{F_\pi}\pd_\mu \pi
-\frac{1}{6F_\pi^3}
\brac{\pi , \brac{\pi , \pd_\mu \pi}}
+2\tr\brac{\Xi (\pi) \cdot \mc{V}_\mu\cdot \Xi^\dagger (\pi) \cdot X_{\perp(3)}^a}X_{\perp(3)}^a
}
\cdot V^{\mu}_{\perp (3)}}
\nn\\
&+\brac{\,\mbox{Terms~for~}(I=0)\, : \mbox{from}~\bar{a}_{(8)} \mbox{to}~\bar{a}_{(14)} }
+\cdots
\label{L_V and L_chi in Appendix}
\ .
}
 Substituting 
the masses and the eigenstates defined in Sec.\,\ref{sec:Vector mesons and pion}
into Eq.\,\eqref{L_V and L_chi in Appendix}, 
we have
\bee{
\mc{L}_V
=&
\tr\brac{\pd_\mu \pi}^2
+
m_{\rho}^2\tr\brac{{\rho}_\mu}^2
+
m_{\rho'}^2\tr\brac{\braa{\rho'}_\mu}^2
+
m_{a_1}^2\tr\brac{\braa{a_1 }_\mu}^2
+
m_{b_1}^2\tr\brac{\braa{b_1 }_\mu}^2
\nn\\
&
+
m_{\omega}^2\tr\brac{{\omega}_\mu}^2
+
m_{\omega'}^2\tr\brac{\braa{\omega'}_\mu}^2
+
m_{f_1}^2\tr\brac{\braa{f_1 }_\mu}^2
+
m_{h_1}^2\tr\brac{\braa{h_1 }_\mu}^2
\nn\\
&
-
\braa{
\frac{1}{3f_\pi^2}
-
\frac{m_{\rho'}^2\sin^2\theta_\rho +m_{\rho}^2\cos^2\theta_\rho }{4g^2f_\pi^4}
}
\frac{4}{i^2}\tr\brac{
\brac{\pi , \pd_\mu \pi}
\brac{\pi , \pd^\mu \pi}
}
+
\braa{
\frac{r_{a_1}m_{a_1}^2}{6gf_\pi^3}}
\frac{4}{i^2}
\tr\brac{
\brac{\pi , \pd_\mu \pi}
\cdot 
\brac{\pi ,\braa{a_1}^{\mu}} 
}
\nn\\
&
+
\braa{\frac{m_\rho^2 \cos \theta_\rho}{\sqrt{2}gf_\pi^2}}
\frac{2\sqrt{2}}{i}
\tr\brac{
\brac{\pi , \pd_\mu \pi}
\cdot 
{\rho}^\mu
}
+
\braa{\frac{m_{\rho'}^2 \sin \theta_\rho}{\sqrt{2}gf_\pi^2}}
\frac{2\sqrt{2}}{i}
\tr\brac{
\brac{\pi , \pd_\mu \pi}
\cdot 
\braa{\rho'}^\mu
}
\nn\\
&
-
\braa{\frac{\sqrt{2}m_{\rho}^2 \cos \theta_\rho}{g}}
2\tr\brac{
\mc{V}_{\mu \parallel}
\cdot {\rho }^\mu
}
-
\braa{\frac{\sqrt{2}m_{\rho'}^2 \sin \theta_\rho}{g}}
2\tr\brac{
\mc{V}_{\mu \parallel}
\cdot \braa{\rho'}^\mu
}
\nn\\
&
-
\braa{\frac{\sqrt{2}m_{\omega}^2 \cos \theta_\omega}{3g_B}}
6\tr\brac{
\mc{V}_{\mu }^0
\cdot {\omega }^\mu
}
-
\braa{\frac{\sqrt{2}m_{\omega'}^2 \sin \theta_\omega}{3g_B}}
6\tr\brac{
\mc{V}_{\mu }^0
\cdot \braa{\omega'}^\mu
}
\nn\\
&
+
f_\pi 2
\tr\brac{
\mc{A}_{\mu }
\braa{\pd^\mu \pi}
}
+
f_\pi 2
\tr\brac{
\mc{A}_{\mu}^0
\braa{\pd^\mu \pi}
}
-
\braa{\frac{r_{a_1}m_{a_1}^2}{g}}
2\tr\brac{
\mc{A}_{\mu }
\cdot \braa{a_1}^{\mu}}
-
\braa{\frac{r_{a_1}m_{a_1}^2}{g}}
2\tr\brac{
\mc{A}_{\mu}^0
\cdot \braa{f_1}^{\mu}}
\nn\\
&
-
\braa{\frac{\sqrt{2}m_{\rho}^2 \cos \theta_\rho}{gf_\pi}}
\frac{2\sqrt{2}}{i} 
\tr\brac{
\mc{A}_{\mu } \cdot \brac{\pi , {\rho }^\mu}
}
-
\braa{\frac{\sqrt{2}m_{\rho'}^2 \sin \theta_\rho}{gf_\pi}}
\frac{2\sqrt{2}}{i} 
\tr\brac{
\mc{A}_{\mu } \cdot \brac{\pi , \braa{\rho' }^\mu}
}
\nn\\
&
-
\braa{ 
\frac{r_{a_1}m_{a_1}^2}{gf_\pi}}
\frac{2\sqrt{2}}{i}\tr\brac{
\mc{V}_{\mu \parallel}
\cdot 
\brac{\pi ,\braa{a_1}^{\mu}}
}
+
\braa{1 -
\frac{m_{\rho'}^2\sin^2\theta_\rho +m_{\rho}^2\cos^2\theta_\rho }{g^2f_\pi^2} }
\frac{2\sqrt{2}}{i}\tr\brac{
\mc{V}_{\mu \parallel}
\cdot \brac{\pi , \pd^\mu \pi}
}
+\cdots \ , 
\label{Lagrangian including alpha}
}
 where the
vector and axial parts of the 
external gauge field are defined as
\bee{
\mc{V}_{\mu \parallel}
=&
\mc{V}_{\mu \parallel}^a
\cdot
S^a
=
\frac{1}{2}\braa{
\mc{R}_{\mu}^a + \mc{L}_{\mu}^a
}
\cdot
S^a
\ ,\nn\\
\mc{A}_{\mu }
=&
\mc{A}_{\mu }^a
\cdot
X_{\perp(3)}^a
=
\frac{1}{2}\braa{
\mc{R}_{\mu}^a - \mc{L}_{\mu}^a
}
\cdot
X_{\perp(3)}^a
}
by using the external fields given in Eq.\,\eqref{definition of external gauge}.
Here Eq.\,\eqref{definition of external gauge} is rewritten as
\bee{
\mc{V}_\mu =
\sqrt{2}\braa{
 \mc{V}_{\mu \parallel}
+
\mc{A}_{\mu }
+
\mc{V}_\mu^0 S^0 +\mc{A}_\mu^0 X_{\perp (3)}^0
}
\ .
}

On the other hand, the kinetic term of the HLS gauge field also gives three point interaction terms:
\bee{
\mc{L}_{\rm int}^{(3)}
=&
-\frac{1}{ig^2}
\tr\brac{\braa{\pd_\mu V_\nu - \pd_\nu V_\mu} \brac{V^\mu, V^\nu}}
\nn\\
=&
-\frac{1}{2\sqrt{2}g^2}
\epsilon^{abc}\braa{\pd_\mu V_{\nu\parallel}^a  - \pd_\nu V_{\mu\parallel}^a}  V^{\mu b}_{\parallel} V^{\nu c}_{\parallel}
-\frac{1}{2\sqrt{2}g^2}
\epsilon^{abc}\braa{\pd_\mu V_{\nu\parallel}^a -\pd_\nu V_{\mu\parallel}^a } V^{\mu b}_{\perp {(i)}} V^{\nu c}_{\perp{(i)}}
\nn\\
&
-\frac{1}{\sqrt{2}g^2}
\epsilon^{ijk}\braa{\pd_\mu V_{\nu\perp {(i)}}^a - \pd_\nu V_{\mu\perp {(i)}}^a } V^{\mu a}_{\perp {(j)}} V^{\nu {(I=0)}}_{\perp{(k)}}
-\frac{1}{\sqrt{2}g^2}
\epsilon^{abc}
\braa{\pd_\mu V_{\nu\perp {(i)}}^a - \pd_\nu V_{\mu\perp {(i)}}^a } V^{\mu b}_{\parallel} V^{\nu c}_{\perp{(i)}}
\nn\\
&
-\frac{1}{2\sqrt{2}g^2}
\epsilon^{ijk}\braa{\pd_\mu V_{\nu\perp {(i)}}^{(I=0)} -\pd_\nu V_{\mu\perp  {(i)}}^{(I=0)} } V^{\mu a}_{\perp {(j)}} V^{\nu a}_{\perp{(k)}}
-\frac{1}{2\sqrt{2}g^2}
\epsilon^{ijk}\braa{\pd_\mu V_{\nu\perp {(i)}}^{(I=0)} -\pd_\nu V_{\mu\perp  {(i)}}^{(I=0)} } V^{\mu {(I=0)}}_{\perp {(j)}} V^{\nu{(I=0)}}_{\perp{(k)}}
}
where $a, b, c = 1,2,3$ corresponding to isospin, $i, j, k=1,2,3$, and
\bee{
V_{\mu}=&
V_{\mu \parallel}
+
V_{\mu \perp (1)}
+
V_{\mu \perp (2)}
+
V_{\mu \perp (3)}
+
V_{\mu \parallel}^{(I=0)}
+
V_{\mu \perp (1)}^{(I=0)}
+
V_{\mu \perp (2)}^{(I=0)}
+
V_{\mu \perp (3)}^{(I=0)}
\ .
}
Since the mixing structures of the gauge fields are given by 
\bee{
V_{\mu \perp{(3)}}^a =& g \braa{a_1}_\mu^a + \frac{r_{a_1}}{ f_\pi}\pd_\mu \pi^a
\ ,~~~~
V_{\mu \parallel}^a =  g\braa{\cos \theta_\rho \rho_\mu^a +\sin \theta_\rho \braa{\rho'}_\mu^a}
\ ,~~~~
V_{\mu \perp{(1)}}^a =  g\braa{\cos \theta_\rho \braa{\rho'}_\mu^a-\sin \theta_\rho \rho_\mu^a }
\ ,
\nn\\
V_{\mu \perp{(1)}}^{(I=0)} =&  g\braa{\cos \theta_\omega \braa{\omega'}_\mu
-\sin \theta_\omega \omega_\mu }
\ ,~~~~
V_{\mu \perp {(2)}}^a  =g\braa{b_1}_\mu^a
\ ,~~~~
V_{\mu \perp {(2)}}^{(I=0)}  = g \braa{h_1}_\mu 
\ ,
}
the three-point  interaction terms related to pion emission are expressed as
\bee{
\mc{L}_{\rm int}^{(3)}
=&
-\frac{1}{2\sqrt{2}g}
\epsilon^{abc}\brac{\cos \theta_\rho  \braa{\pd_\mu  \rho_\nu^a -\pd_\nu  \rho_\mu^a}
+\sin \theta_\rho \braa{\pd_\mu  \braa{\rho'}_\nu^a -\pd_\nu \braa{\rho'}_\mu^a} } 
\braa{g \braa{a_1}^{\mu b} + \frac{r_{a_1}}{ f_\pi}\pd^\mu \pi^b}
\braa{g \braa{a_1}^{\nu c} + \frac{r_{a_1}}{ f_\pi}\pd^\nu \pi^c} 
\nn\\
&
+\frac{1}{\sqrt{2}}
\brac{
\cos \theta_\rho \braa{\pd_\mu  \braa{\rho'}_\nu^a -\pd_\nu \braa{\rho'}_\mu^a}
-\sin \theta_\rho  \braa{\pd_\mu  \rho_\nu^a -\pd_\nu  \rho_\mu^a} } 
\braa{g \braa{a_1}^{\mu a} + \frac{r_{a_1}}{ f_\pi}\pd^\mu \pi^a}
 \braa{h_1}^\nu 
\nn\\
&
-\frac{1}{\sqrt{2}}
\braa{\pd_\mu \braa{b_1}_\nu^a - \pd_\nu \braa{b_1}_\mu^a }
\braa{g \braa{a_1}^{\mu a} + \frac{r_{a_1}}{ f_\pi}\pd^\mu \pi^a}
\braa{\cos \theta_\omega \braa{\omega'}^\nu-\sin \theta_\omega \omega^\nu } 
\nn\\
&
-\frac{1}{\sqrt{2}}
\epsilon^{abc}
\braa{\pd_\mu \braa{a_1}_\nu^a - \pd_\nu \braa{a_1}_\mu^a }
\braa{g \braa{a_1}^{\mu b} + \frac{r_{a_1}}{ f_\pi}\pd^\mu \pi^b}
\braa{\cos \theta_\rho \rho^{\nu c} +\sin \theta_\rho \braa{\rho'}^{\nu c}}
\nn\\
&
-\frac{1}{\sqrt{2}}
\braa{\pd_\mu \braa{h_1}_\nu - \pd_\nu \braa{h_1}_\mu }\braa{g \braa{a_1}^{\mu a} + \frac{r_{a_1}}{ f_\pi}\pd^\mu \pi^a}
\braa{\cos \theta_\rho \braa{\rho'}^{\nu a}-\sin \theta_\rho \rho^{\nu a} }
\nn\\
&
+\frac{1}{\sqrt{2}}
\brac{
\cos \theta_\omega \braa{\pd_\mu  \braa{\omega'}_\nu -\pd_\nu \braa{\omega'}_\mu}
-\sin \theta_\omega  \braa{\pd_\mu  \omega_\nu -\pd_\nu  \omega_\mu} } 
 \braa{g \braa{a_1}^{\mu a} + \frac{r_{a_1}}{ f_\pi}\pd^\mu \pi^a}
 \braa{b_1}^{\nu a} 
+\cdots
\ .
}
These interactions are controlled by only four parameters
\bee{
g
\ ,~~~
r_{a_1}
\ ,~~~
\theta_\rho
\ ,~~~
\theta_\omega
\ ,
}
thanks to the $\mbox{SU(4)}$ symmetry.

\end{widetext}

\section{Intrinsic parity odd terms}
\label{sec:Intrinsic parity odd terms}
Intrinsic parity\,(IP) is a $Z_2$ transformation defined as
\bee{
\braa{\hat{\alpha}_{\mu \parallel}\ ,~ 
\hat{\alpha}_{\mu \parallel}^{(m)}\ ,~ 
\hat{\alpha}_{\mu \perp (1)}^{(m)}}
\ra&
+
\braa{\hat{\alpha}_{\mu \parallel}\ ,~ 
\hat{\alpha}_{\mu \parallel}^{(m)}\ ,~ 
\hat{\alpha}_{\mu \perp (1)}^{(m)}}
\ ,
\nn\\
\braa{
\hat{\alpha}_{\mu \perp (2)}^{(m)}\ ,~
\hat{\alpha}_{\mu \perp (3)}\ ,~
\hat{\alpha}_{\mu \perp (3)}^{(m)}
}
\ra&
- \braa{
\hat{\alpha}_{\mu \perp (2)}^{(m)}\ ,~
\hat{\alpha}_{\mu \perp (3)}\ ,~
\hat{\alpha}_{\mu \perp (3)}^{(m)}
}
\ .
}
The $\mc{O}(p^2)$ terms given in Sec.\,\ref{sec:construction} 
are IP-even.
IP-odd terms are constructed 
at $\mc{O}(p^4)$ of the derivative expansion as
\bee{
\mc{L}^{\rm (IP\,odd)}_{1}=
&
i\epsilon^{\mu\nu\rho\sigma}\tr \brac{
\hat{\alpha}_{\mu\parallel } 
\hat{\alpha}_{\nu \parallel } 
\hat{\alpha}_{\rho\parallel } 
\hat{\alpha}_{\sigma\perp (3)} X^0_{(3)}
}
\ ,
\nn\\
\mc{L}^{\rm (IP\,odd)}_{2}=
&
i\epsilon^{\mu\nu\rho\sigma}\tr \brac{
\hat{\alpha}_{\mu\parallel } 
\hat{\alpha}_{\nu\perp (1) }^{(m)} 
\hat{\alpha}_{\rho \parallel } 
\hat{\alpha}_{\sigma\perp (3)} X^0_{(3)}
\bar{\Sigma} 
}
\ ,
\nn\\
\mc{L}^{\rm (IP\,odd)}_{3}=
&
i\epsilon^{\mu\nu\rho\sigma}\tr \brac{
\hat{\alpha}_{\mu\perp (1) }^{(m)} 
\hat{\alpha}_{\nu\parallel } 
\hat{\alpha}_{\rho\perp (1) }^{(m)} 
\hat{\alpha}_{\sigma\perp (3)} X^0_{(3)}
}
\ ,
\nn\\
\mc{L}^{\rm (IP\,odd)}_{4}=
&
i\epsilon^{\mu\nu\rho\sigma}\tr \brac{
\hat{\alpha}_{\mu\perp (1) }^{(m)} 
\hat{\alpha}_{\nu\perp (1) }^{(m)} 
\hat{\alpha}_{\rho\perp (1) }^{(m)} 
\hat{\alpha}_{\sigma\perp (3)} X^0_{(3)}
\bar{\Sigma} 
}
\ ,
\nn\\
\mc{L}^{\rm (IP\,odd)}_{5}=
&
i\epsilon^{\mu\nu\rho\sigma}\tr \brac{
\hat{\alpha}_{\mu\parallel } 
\hat{\alpha}_{\nu \parallel } 
\hat{\alpha}_{\rho\parallel } 
\hat{\alpha}_{\sigma\perp (3)}^{(m)} X^0_{(3)}
}
\ ,
\nn\\
\mc{L}^{\rm (IP\,odd)}_{6}=
&
i\epsilon^{\mu\nu\rho\sigma}\tr \brac{
\hat{\alpha}_{\mu\parallel } 
\hat{\alpha}_{\nu\perp (1) }^{(m)} 
\hat{\alpha}_{\rho \parallel } 
\hat{\alpha}_{\sigma\perp (3)}^{(m)} X^0_{(3)}
\bar{\Sigma} 
}
\ ,
\nn\\
\mc{L}^{\rm (IP\,odd)}_{7}=
&
i\epsilon^{\mu\nu\rho\sigma}\tr \brac{
\hat{\alpha}_{\mu\perp (1) }^{(m)} 
\hat{\alpha}_{\nu\parallel } 
\hat{\alpha}_{\rho\perp (1) }^{(m)} 
\hat{\alpha}_{\sigma\perp (3)}^{(m)} X^0_{(3)}
}
\ ,
\nn\\
\mc{L}^{\rm (IP\,odd)}_{8}=
&
i\epsilon^{\mu\nu\rho\sigma}\tr \brac{
\hat{\alpha}_{\mu\perp (1) }^{(m)} 
\hat{\alpha}_{\nu\perp (1) }^{(m)} 
\hat{\alpha}_{\rho\perp (1) }^{(m)} 
\hat{\alpha}_{\sigma\perp (3)}^{(m)} X^0_{(3)}
\bar{\Sigma} 
}
\ ,
\nn\\
\mc{L}^{\rm (IP\,odd)}_{9}=
&
i\epsilon^{\mu\nu\rho\sigma}\tr \brac{
\hat{\alpha}_{\mu\parallel } 
\hat{\alpha}_{\nu\perp (3) } 
\hat{\alpha}_{\rho\perp (3) } 
\hat{\alpha}_{\sigma\perp (3)} X^0_{(3)}
}
\ ,
\nn\\
\mc{L}^{\rm (IP\,odd)}_{10}=
&
i\epsilon^{\mu\nu\rho\sigma}\tr \brac{
\hat{\alpha}_{\mu\perp (1) }^{(m)} 
\hat{\alpha}_{\nu\perp (3) } 
\hat{\alpha}_{\rho\perp (3) } 
\hat{\alpha}_{\sigma\perp (3)} X^0_{(3)}
\bar{\Sigma} 
}
\ ,
\nn\\
\mc{L}^{\rm (IP\,odd)}_{11}=
&
i\epsilon^{\mu\nu\rho\sigma}\tr \brac{
\hat{\alpha}_{\mu\parallel } 
\hat{\alpha}_{\nu\perp (3) }  
\hat{\alpha}_{\rho\perp (3) }^{(m)}  
\hat{\alpha}_{\sigma\perp (3)} X^0_{(3)}
}
\ ,
\nn\\
\mc{L}^{\rm (IP\,odd)}_{12}=
&
i\epsilon^{\mu\nu\rho\sigma}\tr \brac{
\hat{\alpha}_{\mu\perp (1) }^{(m)} 
\hat{\alpha}_{\nu\perp (3) } 
\hat{\alpha}_{\rho\perp (3) }^{(m)}  
\hat{\alpha}_{\sigma\perp (3)} X^0_{(3)}
\bar{\Sigma} 
}
\ ,
\nn\\
\mc{L}^{\rm (IP\,odd)}_{13}=
&
i\epsilon^{\mu\nu\rho\sigma}\tr \brac{
\hat{\alpha}_{\mu\parallel } 
\hat{\alpha}_{\nu\perp (3) }^{(m)}  
\hat{\alpha}_{\rho\perp (3) }  
\hat{\alpha}_{\sigma\perp (3)}^{(m)} X^0_{(3)}
}
\ ,
\nn\\
\mc{L}^{\rm (IP\,odd)}_{14}=
&
i\epsilon^{\mu\nu\rho\sigma}\tr \brac{
\hat{\alpha}_{\mu\perp (1) }^{(m)} 
\hat{\alpha}_{\nu\perp (3) }^{(m)} 
\hat{\alpha}_{\rho\perp (3) }  
\hat{\alpha}_{\sigma\perp (3)}^{(m)} X^0_{(3)}
\bar{\Sigma} 
}
\ ,
\nn\\
\mc{L}^{\rm (IP\,odd)}_{15}=
&
i\epsilon^{\mu\nu\rho\sigma}\tr \brac{
\hat{\alpha}_{\mu\parallel } 
\hat{\alpha}_{\nu\perp (3) }^{(m)}  
\hat{\alpha}_{\rho\perp (3) }^{(m)}  
\hat{\alpha}_{\sigma\perp (3)}^{(m)} X^0_{(3)}
}
\ ,
\nn\\
\mc{L}^{\rm (IP\,odd)}_{16}=
&
i\epsilon^{\mu\nu\rho\sigma}\tr \brac{
\hat{\alpha}_{\mu\perp (1) }^{(m)} 
\hat{\alpha}_{\nu\perp (3) }^{(m)} 
\hat{\alpha}_{\rho\perp (3) }^{(m)}  
\hat{\alpha}_{\sigma\perp (3)}^{(m)} X^0_{(3)}
\bar{\Sigma} 
}
\ ,
\nn\\
\mc{L}^{\rm (IP\,odd)}_{17}=
&
i\epsilon^{\mu\nu\rho\sigma}\tr \brac{
\hat{\alpha}_{\mu\parallel  } 
\hat{\alpha}_{\nu\perp (2) }^{(m)} 
\hat{\alpha}_{\rho\perp (3) } 
\hat{\alpha}_{\sigma\perp (2)}^{(m)} X^0_{(3)}
}
\ ,
\nn\\
\mc{L}^{\rm (IP\,odd)}_{18}=
&
i\epsilon^{\mu\nu\rho\sigma}\tr \brac{
\hat{\alpha}_{\mu\perp (1) }^{(m)} 
\hat{\alpha}_{\nu\perp (2) }^{(m)} 
\hat{\alpha}_{\rho\perp (3) } 
\hat{\alpha}_{\sigma\perp (2)}^{(m)} X^0_{(3)}
\bar{\Sigma} 
}
\ ,
\nn\\
\mc{L}^{\rm (IP\,odd)}_{19}=
&
i\epsilon^{\mu\nu\rho\sigma}\tr \brac{
\hat{\alpha}_{\mu\parallel  } 
\hat{\alpha}_{\nu\perp (2) }^{(m)} 
\hat{\alpha}_{\rho\perp (3) }^{(m)} 
\hat{\alpha}_{\sigma\perp (2)}^{(m)} X^0_{(3)}
}
\ ,
\nn\\
\mc{L}^{\rm (IP\,odd)}_{20}=
&
i\epsilon^{\mu\nu\rho\sigma}\tr \brac{
\hat{\alpha}_{\mu\perp (1) }^{(m)} 
\hat{\alpha}_{\nu\perp (2) }^{(m)} 
\hat{\alpha}_{\rho\perp (3) }^{(m)} 
\hat{\alpha}_{\sigma\perp (2)}^{(m)} X^0_{(3)}
\bar{\Sigma} 
}
\ ,
\nn\\
\mc{L}^{\rm (IP\,odd)}_{21}=
&
\epsilon^{\mu\nu\rho\sigma}\tr \brac{
\braa{V_{\mu\nu}^{(m)}}_{\parallel}
\brab{
\hat{\alpha}_{\rho\parallel  } 
, 
\hat{\alpha}_{\sigma\perp (3)}
}
 X^0_{(3)}
}
\ ,
\nn\\
\mc{L}^{\rm (IP\,odd)}_{22}=
&
\epsilon^{\mu\nu\rho\sigma}\tr \brac{
\braa{V_{\mu\nu}^{(m)}}_{\parallel}
\brac{
\hat{\alpha}_{\rho\perp (1) }^{(m)} 
, 
\hat{\alpha}_{\sigma\perp (3)}
}
 X^0_{(3)}
\bar{\Sigma}
}
\ ,
\nn\\
\mc{L}^{\rm (IP\,odd)}_{23}=
&
\epsilon^{\mu\nu\rho\sigma}\tr \brac{
\braa{V_{\mu\nu}^{(m)}}_{\parallel}
\brab{
\hat{\alpha}_{\rho\parallel  } 
, 
\hat{\alpha}_{\sigma\perp (3)}^{(m)}
}
 X^0_{(3)}
}
\ ,
\nn\\
\mc{L}^{\rm (IP\,odd)}_{24}=
&
\epsilon^{\mu\nu\rho\sigma}\tr \brac{
\braa{V_{\mu\nu}^{(m)}}_{\parallel}
\brac{
\hat{\alpha}_{\rho\perp (1) }^{(m)} 
, 
\hat{\alpha}_{\sigma\perp (3)}^{(m)}
}
 X^0_{(3)}
\bar{\Sigma}
}
\ ,
\nn\\
\mc{L}^{\rm (IP\,odd)}_{25}=
&
\epsilon^{\mu\nu\rho\sigma}\tr \brac{
\braa{V_{\mu\nu}^{(m)}}_{\parallel}
\brac{
\hat{\alpha}_{\rho\parallel  } 
, 
\hat{\alpha}_{\sigma\perp (2)}^{(m)}
}
 X^0_{(3)}
\bar{\Sigma} 
}
\ ,
\nn\\
\mc{L}^{\rm (IP\,odd)}_{26}=
&
\epsilon^{\mu\nu\rho\sigma}\tr \brac{
\braa{V_{\mu\nu}^{(m)}}_{\parallel}
\brab{
\hat{\alpha}_{\rho\perp (1) }^{(m)} 
, 
\hat{\alpha}_{\sigma\perp (2)}^{(m)}
}
 X^0_{(3)}
}
\ ,
\nn\\
\mc{L}^{\rm (IP\,odd)}_{27}=
&
\epsilon^{\mu\nu\rho\sigma}\tr \brac{
\braa{V_{\mu\nu}^{(m)}}_{\perp (1)}
\brab{
\hat{\alpha}_{\rho\parallel  } 
, 
\hat{\alpha}_{\sigma\perp (3)}
}
 X^0_{(3)}
\bar{\Sigma} 
}
\ ,
\nn\\
\mc{L}^{\rm (IP\,odd)}_{28}=
&
\epsilon^{\mu\nu\rho\sigma}\tr \brac{
\braa{V_{\mu\nu}^{(m)}}_{\perp (1)}
\brac{
\hat{\alpha}_{\mu\perp (1) }^{(m)} 
, 
\hat{\alpha}_{\sigma\perp (3)}
}
 X^0_{(3)}
}
\ ,
\nn\\
\mc{L}^{\rm (IP\,odd)}_{29}=
&
\epsilon^{\mu\nu\rho\sigma}\tr \brac{
\braa{V_{\mu\nu}^{(m)}}_{\perp (1)}
\brab{
\hat{\alpha}_{\rho\parallel  } 
, 
\hat{\alpha}_{\sigma\perp (3)}^{(m)}
}
 X^0_{(3)}
\bar{\Sigma} 
}
\ ,
\nn\\
\mc{L}^{\rm (IP\,odd)}_{30}=
&
\epsilon^{\mu\nu\rho\sigma}\tr \brac{
\braa{V_{\mu\nu}^{(m)}}_{\perp (1)}
\brac{
\hat{\alpha}_{\mu\perp (1) }^{(m)} 
, 
\hat{\alpha}_{\sigma\perp (3)}^{(m)}
}
 X^0_{(3)}
}
\ ,
\nn\\
\mc{L}^{\rm (IP\,odd)}_{31}=
&
\epsilon^{\mu\nu\rho\sigma}\tr \brac{
\braa{V_{\mu\nu}^{(m)}}_{\perp (1)}
\brac{
\hat{\alpha}_{\rho\parallel  } 
, 
\hat{\alpha}_{\sigma\perp (2)}^{(m)}
}
 X^0_{(3)}
\bar{\Sigma} 
}
\ ,
\nn\\
\mc{L}^{\rm (IP\,odd)}_{32}=
&
\epsilon^{\mu\nu\rho\sigma}\tr \brac{
\braa{V_{\mu\nu}^{(m)}}_{\perp (1)}
\brab{
\hat{\alpha}_{\rho\perp (1) }^{(m)} 
, 
\hat{\alpha}_{\sigma\perp (2)}^{(m)}
}
 X^0_{(3)}
}
\ ,
\nn\\
\mc{L}^{\rm (IP\,odd)}_{33}=
&
\epsilon^{\mu\nu\rho\sigma}\tr \brac{
\braa{V_{\mu\nu}^{(m)}}_{\perp (2)}
\brac{
\hat{\alpha}_{\rho\parallel  } 
, 
\hat{\alpha}_{\sigma\parallel  } 
}
 X^0_{(3)}
\bar{\Sigma} 
}
\ ,
\nn\\
\mc{L}^{\rm (IP\,odd)}_{34}=
&
\epsilon^{\mu\nu\rho\sigma}\tr \brac{
\braa{V_{\mu\nu}^{(m)}}_{\perp (2)}
\brac{
\hat{\alpha}_{\rho\parallel  } 
, 
\hat{\alpha}_{\sigma\perp (1)}^{(m)}
}
 X^0_{(3)}
}
\ ,
\nn\\
\mc{L}^{\rm (IP\,odd)}_{35}=
&
\epsilon^{\mu\nu\rho\sigma}\tr \brac{
\braa{V_{\mu\nu}^{(m)}}_{\perp (2)}
\brac{
\hat{\alpha}_{\rho\perp (1)}^{(m)}
, 
\hat{\alpha}_{\sigma\perp (1)}^{(m)}
}
 X^0_{(3)}
\bar{\Sigma} 
}
\ ,
\nn\\
\mc{L}^{\rm (IP\,odd)}_{36}=
&
\epsilon^{\mu\nu\rho\sigma}\tr \brac{
\braa{V_{\mu\nu}^{(m)}}_{\perp (2)}
\brac{
\hat{\alpha}_{\mu\perp (2) }^{(m)} 
, 
\hat{\alpha}_{\sigma\perp (2)}^{(m)}
}
 X^0_{(3)}
\bar{\Sigma} 
}
\ ,
\nn\\
\mc{L}^{\rm (IP\,odd)}_{37}=
&
\epsilon^{\mu\nu\rho\sigma}\tr \brac{
\braa{V_{\mu\nu}^{(m)}}_{\perp (2)}
\brac{
\hat{\alpha}_{\mu\perp (3) }
, 
\hat{\alpha}_{\sigma\perp (3)}
}
 X^0_{(3)}
\bar{\Sigma} 
}
\ ,
\nn\\
\mc{L}^{\rm (IP\,odd)}_{38}=
&
\epsilon^{\mu\nu\rho\sigma}\tr \brac{
\braa{V_{\mu\nu}^{(m)}}_{\perp (2)}
\brac{
\hat{\alpha}_{\mu\perp (3) }
, 
\hat{\alpha}_{\sigma\perp (3)}^{(m)}
}
 X^0_{(3)}
\bar{\Sigma} 
}
\ ,
\nn\\
\mc{L}^{\rm (IP\,odd)}_{39}=
&
\epsilon^{\mu\nu\rho\sigma}\tr \brac{
\braa{V_{\mu\nu}^{(m)}}_{\perp (2)}
\brac{
\hat{\alpha}_{\mu\perp (3) }^{(m)}
, 
\hat{\alpha}_{\sigma\perp (3)}^{(m)}
}
 X^0_{(3)}
\bar{\Sigma} 
}
\ ,
\nn\\
\mc{L}^{\rm (IP\,odd)}_{40}=
&
\epsilon^{\mu\nu\rho\sigma}\tr \brac{
\braa{V_{\mu\nu}^{(m)}}_{\perp (3)}
\brab{
\hat{\alpha}_{\rho\parallel  } 
, 
\hat{\alpha}_{\sigma\parallel  } 
}
 X^0_{(3)}
}
\ ,
\nn\\
\mc{L}^{\rm (IP\,odd)}_{41}=
&
\epsilon^{\mu\nu\rho\sigma}\tr \brac{
\braa{V_{\mu\nu}^{(m)}}_{\perp (3)}
\brab{
\hat{\alpha}_{\rho\parallel  } 
, 
\hat{\alpha}_{\sigma\perp (1)}^{(m)}
}
 X^0_{(3)}
\bar{\Sigma} 
}
\ ,
\nn\\
\mc{L}^{\rm (IP\,odd)}_{42}=
&
\epsilon^{\mu\nu\rho\sigma}\tr \brac{
\braa{V_{\mu\nu}^{(m)}}_{\perp (3)}
\brab{
\hat{\alpha}_{\rho\perp (1)}^{(m)}
, 
\hat{\alpha}_{\sigma\perp (1)}^{(m)}
}
 X^0_{(3)}
}
\ ,
\nn\\
\mc{L}^{\rm (IP\,odd)}_{43}=
&
\epsilon^{\mu\nu\rho\sigma}\tr \brac{
\braa{V_{\mu\nu}^{(m)}}_{\perp (3)}
\brab{
\hat{\alpha}_{\mu\perp (2) }^{(m)} 
, 
\hat{\alpha}_{\sigma\perp (2)}^{(m)}
}
 X^0_{(3)}
}
\ ,
\nn\\
\mc{L}^{\rm (IP\,odd)}_{44}=
&
\epsilon^{\mu\nu\rho\sigma}\tr \brac{
\braa{V_{\mu\nu}^{(m)}}_{\perp (3)}
\brab{
\hat{\alpha}_{\mu\perp (3) }
, 
\hat{\alpha}_{\sigma\perp (3)}
}
 X^0_{(3)}
}
\ ,
\nn\\
\mc{L}^{\rm (IP\,odd)}_{45}=
&
\epsilon^{\mu\nu\rho\sigma}\tr \brac{
\braa{V_{\mu\nu}^{(m)}}_{\perp (2)}
\brab{
\hat{\alpha}_{\mu\perp (3) }
, 
\hat{\alpha}_{\sigma\perp (3)}^{(m)}
}
 X^0_{(3)}
}
\ ,
\nn\\
\mc{L}^{\rm (IP\,odd)}_{46}=
&
\epsilon^{\mu\nu\rho\sigma}\tr \brac{
\braa{V_{\mu\nu}^{(m)}}_{\perp (2)}
\brab{
\hat{\alpha}_{\mu\perp (3) }^{(m)}
, 
\hat{\alpha}_{\sigma\perp (3)}^{(m)}
}
 X^0_{(3)}
}
\ ,
\nn\\
\mc{L}^{\rm (IP\,odd)}_{47}=
&
i\epsilon^{\mu\nu\rho\sigma}\tr \brac{
\braa{V_{\mu\nu}^{(m)}}_{\parallel}
\braa{V_{\rho\sigma}^{(m)}}_{\perp (2)}
 X^0_{(3)}
\bar{\Sigma}
}
\ ,
\nn\\
\mc{L}^{\rm (IP\,odd)}_{48}=
&
i\epsilon^{\mu\nu\rho\sigma}\tr \brac{
\braa{V_{\mu\nu}^{(m)}}_{\perp (1)}
\braa{V_{\rho\sigma}^{(m)}}_{\perp (2)}
 X^0_{(3)}
}
\ ,
\nn\\
\cdots
\ ,
\label{IP odd terms}
}
where we required the invariance under the chiral transformation as well as $\mc{P}$ and $\mc{C}$.
The operators explicitly given in Eq.\,\eqref{IP odd terms} include interactions among $\pi$, $\rho$, $\rho'$, $a_1$, and $b_1$, which are the iso-triplet.
The symbol ``$\cdots$'' in Eq.\,\eqref{IP odd terms} expresses that operators including the iso-singlet variables such as $\hat{\alpha}_{\mu\parallel}^{(I=0)}$ are also allowed.
For convenience, we used the field strength $\braa{V_{\mu\nu}^{(m)}}$ defined as
\bee{
\braa{V_{\mu\nu}^{(m)}}_{\parallel}
\equiv&
2\tr\brac{\Xi_{m} \cdot V_{\mu\nu}
\cdot \Xi_{m}^\dagger \cdot S^a
}
S^a
\ ,
\nn\\
\braa{V_{\mu\nu}^{(m)}}_{\perp (1)}
\equiv&
2\tr\brac{\Xi_{m} \cdot V_{\mu\nu}\cdot \Xi_{m}^\dagger  \cdot X_{(1)}^a}
X_{(1)}^a
\ ,
\nn\\
\braa{V_{\mu\nu}^{(m)}}_{\perp (2)}
\equiv&
2\tr\brac{\Xi_{m} \cdot V_{\mu\nu}\cdot \Xi_{m}^\dagger  \cdot X_{(2)}^a}
X_{(2)}^a
\ ,
\nn\\
\braa{V_{\mu\nu}^{(m)}}_{\perp (3)}
\equiv&
2\tr\brac{\Xi_{m} \cdot V_{\mu\nu}\cdot \Xi_{m}^\dagger  \cdot X_{(3)}^a}
X_{(3)}^a
\ ,
\nn\\
\braa{V_{\mu\nu}^{(m) (I=0)}}_{\parallel}
\equiv&
2\tr\brac{\Xi_{m} \cdot V_{\mu\nu}
\cdot \Xi_{m}^\dagger \cdot S^0
}
S^0
\ ,
\nn\\
\braa{V_{\mu\nu}^{(m) (I=0)}}_{\perp (1)}
\equiv&
2\tr\brac{\Xi_{m} \cdot V_{\mu\nu}\cdot \Xi_{m}^\dagger  \cdot X_{(1)}^0}
X_{(1)}^0
\ ,
\nn\\
\braa{V_{\mu\nu}^{(m) (I=0)}}_{\perp (2)}
\equiv&
2\tr\brac{\Xi_{m} \cdot V_{\mu\nu}\cdot \Xi_{m}^\dagger  \cdot X_{(2)}^0}
X_{(2)}^0
\ ,
\nn\\
\braa{V_{\mu\nu}^{(m) (I=0)}}_{\perp (3)}
\equiv&
2\tr\brac{\Xi_{m} \cdot V_{\mu\nu}\cdot \Xi_{m}^\dagger  \cdot X_{(3)}^0}
X_{(3)}^0
\ ,
}
We can list the  operators including 
$\hat{\alpha}_{\mu \parallel }^{(m)} $,
$\hat{\alpha}_{\mu \parallel }^{(m) (I=0)}$, or external gauge field $\mc{V}_{\mu\nu}$
 in a similar way.

\section{Decays to $\pi + \gamma$}
\label{sec:to pi gamma}

In this appendix, we calculate the decay widths of the spin-1 mesons to $\pi$ and $\gamma$. 

First we calculate the decay width of  the $a_1 \ra \pi \gamma$. 
The relevant $a_1\pi\gamma$, $a_1\rho\pi$ and $a_1\rho'\pi$  vertex functions  are written as
\bee{
&\Gamma^{(\mu a) (\nu) (c)}({a_1} , \gamma ,\pi)
=
-e g_{a_1 \mc{V} \pi}
\epsilon^{a3c}g^{\mu\nu}
\ ,
\nn\\
&\Gamma^{(\mu a) (\nu b) (c)}(\braa{p_{a_1}} , \braa{p_\rho} , p_\pi)
\nn\\
=&
g_{\rho a_1 \pi}
\epsilon^{abc}
\brac{
\braa{p_{a_1}}^2 P^{\mu \nu} \braa{p_{a_1}}
-
\braa{p_\rho}^2 P^{\mu \nu} \braa{p_\rho}
}
\ ,
\nn\\
&\Gamma^{(\mu a) (\nu b) (c)}(\braa{p_{a_1}} , \braa{p_{\rho'}} , p_\pi)
\nn\\
=&
g_{\rho' a_1 \pi}
\epsilon^{abc}
\brac{
\braa{p_{a_1}}^2 P^{\mu \nu} \braa{p_{a_1}}
-
\braa{p_{\rho'}}^2 P^{\mu \nu} \braa{p_{\rho'}}
}
}
where $g_{a_1 \mc{V} \pi}\equiv\frac{m_{a_1}^2r_{a_1}}{ gf_\pi}$.
By using these effective vertex functions, 
we can 
calculate 
 the 
amplitude of $a_1 \ra \pi \gamma$; 
\bee{
&
\mc{M} \braa{ a_1 (p_{a_1})^a_\mu \ra \pi (p_{\pi})^b \gamma (p_{\gamma} )_\nu}
\nn\\
=&
e
\epsilon^{ab3}
\epsilon_\nu (p_\gamma )^*
\epsilon_\mu (p_{a_1})
\nn\\
&\times
g^{\mu\nu}\braa{
-g_{a_1 \mc{V} \pi}
+ \frac{g_{\rho}}{m_{\rho}^2}g_{\rho a_1 \pi}
+ \frac{g_{\rho'}}{m_{\rho'}^2}g_{\rho' a_1 \pi}
}
}
where  $\epsilon^\mu (p)$ is the polarization vector and 
we used $p_\gamma^2 =0$.  
 Then, by using  Eqs.\,\eqref{GT relation} and \eqref{coupling grho}, 
 the above amplitude vanishes: 
${\mathcal M} \braa{ a_1 (p_{a_1})^a_\mu \ra \pi (p_{\pi})^b \gamma (p_{\gamma} )_\nu} = 0$. 
This implies that the decay width for $a_1 \ra \pi\gamma$ vanishes;
\bee{
&
\Gamma (a_1 \ra \pi\gamma)
=
\frac{\brad{\vec{p}_{\pi}}}{8\pi m_{a_1}^2}
\frac{1}{9}
\sum \brad{\mc{M}}^2
=0
\ .
\label{Decay width a1 to pigamma}
}
The model with the generalized HLS~\cite{Bando:1987ym} also shows that the partial width  of  $a_1 \ra \pi\gamma$ vanishes  at the $\mc{O}(p^2)$ order.

Since the mass of the initial particle is around $1 \mbox{GeV}$ while $m_\pi\simeq 137 \,\mbox{MeV}$ and $m_\gamma=0$,
the momentum of the outgoing particles is about $1 \mbox{GeV}$.
This means that the higher order contribution of the derivative expansion  may  not be small.
The $\mc{O}(p^4)$ contribution is estimated as $\mc{M}^{\mc{O}(p^4)} \sim 100 \mbox{MeV}$
because of $\braa{\frac{\brad{\vec{p}_\pi}}{4\pi f_\pi}}^2 \simeq\braa{\frac{m_{a_1}/2}{4\pi f_\pi}}^2 \simeq 0.3$ and $\brad{\frac{e m_{a_1}^2r_{a_1}}{ gf_\pi}}\sim 330 \mbox{MeV}$.
One can find that $\Gamma (a_1 \ra \pi\gamma) $ is of order $100 \mbox{keV}$.
Therefore, the deviation between our result and the experiments, 
$\Gamma(a_1 \ra \pi\gamma)=640\pm 246 \mbox{keV}$, 
is understood as the contribution of the $\mc{O}(p^4)$ order.

Similarly to the $a_1 \to \pi \gamma$ decay, the 
$b_1 \ra \pi \gamma$ and $h_1 \ra \pi \gamma$ amplitudes vanish at the $\mc{O} (p^2)$ order, respectively;
\bee{
&\mc{M} (\braa{b_1}^a \ra \pi^b \gamma)
\nn\\
=&
e
\delta^{ab}
\epsilon_\nu (p_\gamma)^*
\epsilon_\mu (p_{b_1})
g^{\mu\nu}\braa{
\frac{g_{\omega}}{m_{\omega}^2}g_{\omega b_1 \pi}
+ \frac{g_{\omega'}}{m_{\omega'}^2}g_{\omega' b_1 \pi}
}
=0
\ ,
\nn\\
&
\mc{M} (h_1 \ra \pi^0 \gamma)
\nn\\
=&
e
\epsilon_\nu (p_\gamma)^*
\epsilon_\mu (p_{h_1})
g^{\mu\nu}\braa{
\frac{g_{\rho}}{m_{\rho}^2}g_{\rho h_1 \pi}
+ \frac{g_{\rho'}}{m_{\rho'}^2}g_{\rho' h_1 \pi}
}
=0
}
where we used 
$p_\gamma^2=0$ and Eqs.\,\eqref{GT relation}, \eqref{coupling grho}, and \eqref{coupling gomega}. 
Then, their decay widths are
\bee{
\Gamma (b_1 \ra \pi \gamma) 
=
\Gamma (h_1 \ra \pi \gamma) 
=0
\label{Decay width b1 to pigamma}
\ .
}
The empirical value $\Gamma(b_1 \ra \pi\gamma)=230\pm 60 \mbox{keV}$ is the same order as the $\mc{O}(p^4)$ contribution estimated for the decay of the $a_1$ meson.


\begin{thebibliography}{100}


\bibitem{Agashe:2014kda} 
  K.~A.~Olive {\it et al.} [Particle Data Group Collaboration],
  Chin.\ Phys.\ C {\bf 38}, 090001 (2014).


\bibitem{Denissenya:2014ywa} 
  M.~Denissenya, L.~Y.~Glozman and C.~B.~Lang,
  Phys.\ Rev.\ D {\bf 91}, 034505 (2015).


\bibitem{Glozman:2015qva} 
  L.~Y.~Glozman and M.~Pak,
  Phys.\ Rev.\ D {\bf 92}, 016001 (2015).


\bibitem{Cohen:2015ekf} 
  T.~D.~Cohen,
  Phys.\ Rev.\ D {\bf 93}, 034508 (2016).


\bibitem{Shifman:2016efc} 
  M.~Shifman,
  Phys.\ Rev.\ D {\bf 93}, 074035 (2016).


\bibitem{Bando:1984ej} 
  M.~Bando, T.~Kugo, S.~Uehara, K.~Yamawaki and T.~Yanagida,
  Phys.\ Rev.\ Lett.\  {\bf 54}, 1215 (1985).





\bibitem{Bando:1984pw} 
  M.~Bando, T.~Kugo and K.~Yamawaki,
  Prog.\ Theor.\ Phys.\  {\bf 73}, 1541 (1985).

\bibitem{Bando:1985rf} 
  M.~Bando, T.~Kugo and K.~Yamawaki,
  Nucl.\ Phys.\ B {\bf 259}, 493 (1985).

\bibitem{Bando:1987br} 
  M.~Bando, T.~Kugo and K.~Yamawaki,
  Phys.\ Rept.\  {\bf 164}, 217 (1988).

\bibitem{Bando:1987ym} 
  M.~Bando, T.~Fujiwara and K.~Yamawaki,
  Prog.\ Theor.\ Phys.\  {\bf 79}, 1140 (1988).

\bibitem{Harada:2003jx} 
  M.~Harada and K.~Yamawaki,
  Phys.\ Rept.\  {\bf 381}, 1 (2003).


\bibitem{Amendolia:1986wj} 
  S.~R.~Amendolia {\it et al.} [NA7 Collaboration],
  Nucl.\ Phys.\ B {\bf 277}, 168 (1986).

\bibitem{Bebek:1977pe} 
  C.~J.~Bebek {\it et al.},
  Phys.\ Rev.\ D {\bf 17}, 1693 (1978).

\bibitem{Volmer:2000ek} 
  J.~Volmer {\it et al.} [Jefferson Lab F(pi) Collaboration],
  Phys.\ Rev.\ Lett.\  {\bf 86}, 1713 (2001).

\bibitem{Horn:2006tm} 
  T.~Horn {\it et al.} [Jefferson Lab F(pi)-2 Collaboration],
  Phys.\ Rev.\ Lett.\  {\bf 97}, 192001 (2006).

\bibitem{Tadevosyan:2007yd} 
  V.~Tadevosyan {\it et al.} [Jefferson Lab F(pi) Collaboration],
  Phys.\ Rev.\ C {\bf 75}, 055205 (2007).

\bibitem{Harada:2010cn} 
  M.~Harada, S.~Matsuzaki and K.~Yamawaki,
  Phys.\ Rev.\ D {\bf 82}, 076010 (2010).

\bibitem{Nagahiro:2008cv} 
  H.~Nagahiro, L.~Roca, A.~Hosaka and E.~Oset,
  Phys.\ Rev.\ D {\bf 79}, 014015 (2009).


\end{thebibliography}
\end{document}